\documentclass[12pt]{article}

\usepackage{mathtools, nccmath}
\usepackage{verbatim}
\usepackage{amsfonts}
\usepackage{amssymb}
\usepackage{amsmath}
\usepackage[force]{feynmp-auto}
\usepackage{braket}
\usepackage{dsfont}
\usepackage{authblk}
\usepackage[margin=1.3cm]{caption}
\usepackage{blkarray}
\usepackage{graphicx}
\usepackage{caption}
\usepackage{subcaption}
\usepackage{slashed}
\usepackage{color,soul}
\usepackage{bm}
\usepackage{xcolor}
\usepackage{microtype}

\usepackage{jheppub}

\numberwithin{equation}{section}

\newcommand{\hd}{\hat{\partial}}
\newcommand{\wD}{\hat{\nabla}^H}
\newcommand{\mL}{\mathcal{L}}
\newcommand{\mO}{\mathcal{O}}

\newcommand{\pd}{\partial}

\newcommand{\mM}{\mathcal{M}}	
\newcommand{\vp}{\bm{p}}
\newcommand{\vk}{\bm{k}}
\newcommand{\vpp}{\bm{p}^{2}}
\newcommand{\vkk}{\bm{k}^{2}}

\preprint{DESY 20-086}

\begin{document}

\author[1]{G. Cuomo,}
\author[1]{A. Esposito,}
\author[1,2,3]{E. Gendy,}
\author[1]{A. Khmelnitsky,}
\author[1]{A. Monin,}
\author[1]{R. Rattazzi}

\affiliation[1]{Theoretical Particle Physics Laboratory (LPTP), Institute of Physics, EPFL, 1015 Lausanne, Switzerlands}
\affiliation[2]{DESY, Notkestra{\ss}e 85, D-22607 Hamburg, Germany}
\affiliation[3]{Institute of Theoretical Physics, Universit\"at Hamburg, 22761 Hamburg, Germany}

\emailAdd{gabriel.cuomo@epfl.ch}
\emailAdd{angelo.esposito@epfl.ch}
\emailAdd{emanuele.gendy@desy.de}
\emailAdd{andrey.khmelnitskiy@epfl.ch}
\emailAdd{alexander.monin@epfl.ch}
\emailAdd{riccardo.rattazzi@epfl.ch}

\abstract{At finite density, the spontaneous  breakdown of an internal non-Abelian symmetry dictates, along with gapless modes,  modes whose gap is fixed by the algebra and proportional to the chemical potential: the gapped Goldstones. Generically the gap of these states is comparable to that of other non-universal excitations or to the energy scale where the dynamics is strongly coupled. This makes it  non-straightforward to derive a universal effective field theory (EFT)  description realizing all the symmetries.
Focusing on the illustrative example of a fully broken $SU(2)$ group, we demonstrate that such an EFT can be constructed by {carving out around} the Goldstones, gapless and gapped, at small 3-momentum. The rules governing the EFT, where the  gapless Goldstones are  {\it soft } while the gapped ones are {\it slow}, are {those of standard nonrelativistic EFTs,
like for instance nonrelativistic QED}. In particular, the EFT Lagrangian formally preserves gapped Goldstone number, and processes where  such number is not conserved are described inclusively by allowing for imaginary parts in the Wilson coefficients.  Thus, while the symmetry is manifestly realized in the EFT, unitarity is not.
We comment on the application of our construction to the study of the large charge sector of conformal field theories with non-Abelian symmetries.}

%


\title{Gapped Goldstones at the cut-off scale: a non-relativistic EFT}

\maketitle

\section{Introduction}

{Spontaneously broken symmetries have far reaching consequences in the study of physical systems. That is  mainly because of the existence of Nambu-Goldstone bosons~\cite{Goldstone,Nambu}, whose low-energy dynamics is largely dictated by symmetry, independently of other details of the microscopic physics ~\cite{CCWZ1,CCWZ2,Weinberg2}.
As a result, the experimental study of the dynamics of Goldstone bosons at low energies and long distances allows to robustly infer the nature of fundamental symmetries and the pattern of their spontaneous breaking.}


In a standard Lorentz invariant setup there are as many Goldstones as broken generators, they are all massless and move at the speed of light. 
However, Nature is pervaded with systems that spontaneously break spacetime symmetries as well, in which case Goldstone theorem allows for a much richer set of possibilities (see, e.g.,~\cite{LangeNRGoldstones,NielsenNRGoldstones,WatanabeRedundancies,BraunerNonRelNGB}). In this work we focus on those systems that are at finite density for a certain spontaneously broken charge. When the latter does not commute with other broken charges, the spectrum of the theory contains  the so-called \emph{gapped Goldstones}~\cite{Morchio:1987aw,Strocchi:2008gsa, Nicolis_Theorem,Nicolis_More,WatanabeMNGB}.

More precisely, consider a relativistic system that is at finite density for a given charge $Q$ and whose time evolution is governed by a Hamiltonian $H$. In this case, the ground state of the system can be found as the state with lowest eigenvalue with respect to the modified Hamiltonian~(see, for instance,~\cite{Nicolis_SSP})
\begin{align}
\bar H = H + \mu Q \, ,
\end{align}
where $\mu$ is the chemical potential.
In this work we focus on systems of this sort that break boost invariance (like all condensed matter states~\cite{Nicolis_Zoology}), time translations generated by $H$, the internal charge $Q$, as well as another set of internal charges $Q_i$. The modified Hamiltonian $\bar H$ is unbroken by construction. When $Q$ does not commute with some of the $Q_i$'s, Goldstone theorem implies the existence of both gapless modes and gapped ones,\footnote{Strictly speaking, nonrelativistic Goldstone theorem requires the existence of zero-momentum excitations, but does not say anything about finite momentum ones. For instance, phonons in superfluids have a finite width, which vanishes in the limit where their momentum goes to zero (see, e.g.,~\cite{Maris:1977zz}).} whose gap, $\omega(\bm{k}=0)\propto\mu$, is completely fixed nonperturbatively~\cite{Morchio:1987aw,Nicolis_Theorem}. Independently of the presence of the gap, all Goldstone modes share a defining property: their scattering amplitudes vanish  with their 3-momentum---the so-called Adler's zeros~\cite{Brauner}.\footnote{Note that the presence of Adler's zeros for gapless Goldstones is not always guaranteed due to possible kinematic singularities, cf.~\cite{Brauner}. On the other hand, the gap of the gapped Goldstones precludes these singularities, and Adler's zeros for them are always present.} {In other words, all Goldstone bosons are free when their 3-momentum vanishes. An effective field theory (EFT) description of their dynamics should then focus on the regime of small 3-momentum. For  gapless modes this coincides with the regime of low {\it energy}, while for the gapped ones it instead coincides with the regime of low {\it  kinetic energy} or, equivalently, low velocity.}

The presence of both {gapless 
and  gapped modes, however},
makes the piecing together of an EFT approach not  straightforward. This is immediately appreciated by considering the process of annihilation of two gapped modes into two gapless ones; a process that is generically allowed. Even if the {spatial momentum} of the incoming states approaches zero, their total energy is of order $\mu$, and so are the momenta of the final state quanta. Now, when the underlying microscopic dynamics is strong, the gap scale $\mu$ should
coincide, by simple dimensional analysis, with the momentum scale where the gapless modes become strongly coupled.\footnote{That is, for instance, the case in QCD, where the $\rho$ mass parametrically coincides  with the scale where $\pi$ interactions become strong}  In that case, while {the amplitude is still  suppressed at small initial  momenta}, the emission and exchange of additional gapless modes will {contribute  $\mathcal{O}(1)$} relative corrections to the total rate, thus making it practically incalculable. In other words the interaction among slow gapped modes can lead to the production of very energetic gapless ones, beyond the reach of the ordinary EFT description of their dynamics. 

The question is then how to properly describe this state of affairs. On the one hand, the gapped Goldstones are free at zero {momentum/velocity}, as dictated by symmetry, while on the other, at arbitrarily small velocity, the processes involving them do not seem calculable. Integrating out the gapped modes in favor of an ordinary EFT for the gapless ones, while certainly doable, does not seem satisfactory, {as it  would preclude describing those aspects  of the dynamics that are dictated by symmetry (like the relation between the gap and the chemical potential or the freedom of gapped modes at zero velocity).} Relatedly that would make the underlying symmetry breaking pattern not visible in the EFT.\footnote{For instance in the case of a fully broken non-Abelian group $G$ the gapless modes are purely described by the spontaneous breaking of the  Cartan subgroup of $G$ \cite{MoninCFT}, with  seemingly no visible low-energy remnant of the non-Abelian nature of the original group.} 
In this paper we address the problem by constructing  a proper EFT  that allows for a  more limited but systematic description of the gapped Goldstone dynamics. The construction is fully analogous to the  nonrelativistic EFT (NREFT) used, for instance, to describe positronium~\cite{NRQED}. Like in the positronium case, the price to pay is the existence of absorbitive (imaginary) terms in the effective action~\cite{Labelle,Braaten}. Within this NREFT approach, we shall  illustrate how to describe the dynamics in a systematic small momentum expansion.

Besides the above mentioned conceptual issues,
understanding the consequences of a spontaneously broken non-Abelian symmetry at finite density is also  a question of phenomenological relevance. Indeed, gapped Goldstones appear in many different contexts~\cite{WatanabeMNGB}, ranging from condensed matter systems~\cite{Kohn,Leutwyler,SpinResonance}, to QCD at finite isospin density in the chiral limit~\cite{Kaplan_KaonCondensate,Son_Kaon1,Son_Kaon2,Brown_NeutronStar}. Furthermore, they are also relevant  in conformal field theories, where one can use the state/operator correspondence to map operators with large internal quantum numbers to finite density states~\cite{Hellerman,MoninCFT,Bern1,HellermanO41,BootstrapLargeQ}. As such, gapped Goldstones appear in the description of  the spectrum of deformations of critical points in statistical physics.

\vspace{1em}

In this paper {we illustrate} our ideas by focusing on a simple system with an $SU(2)$ symmetry fully broken by the finite density of one of its charges. The resulting spectrum  features a gapless and a gapped Goldstone, whose gap is precisely $\mu$. In section~\ref{SecMassiveNGBreview} we introduce a simple model that exhibits this symmetry breaking pattern and verify the presence of Adler's zero in the amplitudes for the gapped Goldstones. This will be our benchmark for the rest of the paper. 
{In section~\ref{SecNREFT} we construct a nonrelativistic effective field theory for gapless and gapped Goldstones at small 3-momentum, showing how their interactions are constrained by the full symmetry group. Remarkably, such a construction is applicable for any value of the chemical potential, even when it is of the same order as the UV cutoff of the theory.}
{In order to account for the gapped Goldstone's decay or annihilation, we argue that} the NREFT must contain imaginary coefficients, {which makes it  non-unitary. The lack of unitarity is  simply due to the limited class of degrees of freedom that make up our EFT, and is of course not a fundamental property.} Power counting and interactions in such a theory are analyzed in detail. 
Finally, in section~\ref{secSO(3)=U(1)} we discuss the reasons why there is no remnant of the non-Abelian part of the broken symmetry at energies much smaller than the chemical potential. {In the Conclusions we comment on possible applications of this NREFT, with particular attention to the case of a strongly interacting conformal $O(3)$ model. }

\section{A benchmark model: the linear triplet}\label{SecMassiveNGBreview}

{In this section we present a simple model with  internal $SU(2)$ symmetry, admitting a  finite density state for one of the charges where $SU(2)$ and time translations are broken down to a diagonal subgroup, $H\times SU(2)\to \bar H$. We study perturbations around such state, identify the gapped Goldstone modes and examine the amplitudes for their scattering and annihilation in the regime where their {3-momentum} is small.}
The model is weakly coupled and renormalizable, and hence all observables can be computed perturbatively.
Because of that, we will use it as the main example to match the effective theory developed in the rest of the paper.

\subsection{The model} \label{sec:themodel}

Consider the following renormalizable Lagrangian for an $O(3)$ triplet $\bm\Phi$ in four spacetime dimensions:
\begin{align}\label{eqTripletLagrangian}
\mL=\frac{1}{2}(\pd\bm\Phi)^2-\frac{m^2}{2}
\bm\Phi^2-\frac{\lambda}{4}\bm\Phi^4\,,
\end{align}
where $\lambda>0$, and we do not make any assumptions on the sign on $m^2$. The classical field configuration that realizes the desired symmetry breaking pattern is
\begin{align} \label{eqTripletVEV}
\bm\Phi_0=e^{-i\mu t Q_3}
\begin{pmatrix} \phi_0 \\ 0 \\ 0 \end{pmatrix}\,,\qquad\phi_0^2=\frac{\mu^2-m^2}{\lambda} > 0 \,,
\end{align}
where $(Q_i)_{jk}=-i\epsilon_{ijk}$ are the generators in the defining representation of $SO(3)$. If $m^2>0$ then spontaneous symmetry breaking happens only for $\mu^2 > m^2$. The state described by this configuration is indeed at finite density for the charge $Q_3$, as one can check by computing the corresponding Noether's current. Moreover, since it depends explicitly on time, this vacuum expectation value {(VEV)} breaks both boosts and time translations.\footnote{{Note that the VEV~\eqref{eqTripletVEV} also breaks Galilei boosts in the non-relativistic limit. This can be seen in different ways. Most simply, since it singles out a particular reference frame, boosts must be broken regardless on whether one is considering Lorentz or Galilei. Equivalently, one can notice that both groups feature the same number of charges, while this theory preserves a smaller number of them. Or, more explicitly, we have that, under a boost with velocity $\bm v$, the phase transforms as $\psi \to \psi + t\bm{v} \cdot\bm{\nabla}\psi - \frac{\bm{v}\cdot\bm{x}}{c^2}\partial_0\psi + \frac{1}{2}\frac{v^2}{c^2}t\partial_0\psi + \mO(1/c^4)$, and hence the chemical potential ($\mu\sim m c^2$~\cite{Nicolis:2017eqo}) transforms under non-relativistic boosts ($c\to\infty$) as $\mu t \to \mu t - m\bm{v}\cdot\bm{x}+\frac{1}{2}mv^2t$, making the VEV not invariant.}}
On top of that, it also breaks the internal $O(3)$ symmetry down to $\mathds{Z}_2$ corresponding to $\Phi_3 \to - \Phi_3$, but preserves the combination $\bar H = H + \mu Q_3$. This is then precisely {a setup where} the charge at finite density does not commute with other broken charges.

Before studying the full spectrum, let us give a simple {argument for} the existence of gapped Goldstones. Starting from the configuration~\eqref{eqTripletVEV} and performing, say, a {small rotation along $Q_1$, one obtains another solution, where the third component of the triplet oscillates with precisely frequency $\mu$: $\delta\Phi_3(x)=-\phi_0 \sin\mu t$}. The existence of a mode {with energy $\mu$ when at rest is therefore dictated by  $SU(2)$.} This is parallel to what happens with a rotation generated by $Q_3$, which instead ensures the existence of a gapless mode. {At the same time, this {provides an intuitive argument for} why gapped Goldstones are free when they are at rest: their zero-mode corresponds to nothing but a global transformation.}\footnote{Note that, at infinite volume, the action of the global charges is not defined, and consequently neither is the zero-mode. Strictly speaking one should work at finite volume, 
{and add an infinitesimal perturbation explicitly breaking the symmetry before taking the infinite volume limit \cite{BraunerNonRelNGB,Weinberg2}.}}


The fluctuations around equilibrium can be conveniently parametrized in terms of three real fields, $\psi(x)$, $\theta(x)$ and $h(x)$:
\begin{align} \label{eqTripletFluctuations}
\bm\Phi(x)=e^{-i\left(\mu t+\psi(x)/\phi_0\right) Q_3}
\begin{pmatrix} \phi_0+h(x) \\ 0 \\ \theta(x) \end{pmatrix}\,.
\end{align}
The unbroken $\mathds{Z}_2$ acts as $\theta\to-\theta$. The Lagrangian then reads
\begin{align}
\mL&=\frac{1}{2} (\pd \theta)^2 
-\frac{\mu^2}{2}\theta^2
+\frac12(\pd\psi)^2
+\frac{1}{2}(\pd h)^2
+2\mu h\dot{\psi}
-\lambda\phi_0^2 h^2 \label{eqTripletLagrangianQuadratic} \\
&\quad -\lambda\phi_0h\left( h^2+ \theta^2\right)
-\frac{\lambda}{4}\left(h^4+\theta^4+2h^2\theta^2\right)
+\frac{\mu}{\phi_0}h^2\dot{\psi}
+\frac{1 }{\phi_0} h (\pd\psi)^2 +\frac{1}{2\phi_0^2}h^2(\pd\psi)^2\,. \notag
\end{align}
Extracting the propagator from \eqref{eqTripletLagrangianQuadratic}, one finds that, as expected, the spectrum of the theory consists of
\begin{itemize}
\item A gapless Goldstone, $\pi_3$, with dispersion relation
\begin{align}
\omega^2_k&=k^2+3\mu^2-m^2-\sqrt{(3\mu^2-m^2)^2+4k^2\mu^2} \notag 
\\&=\frac{\mu^2-m^2}{3\mu^2-m^2}k^2+\mO\left(\frac{k^4}{\mu^2}\right)\,. \label{dispi3}
\end{align}
\item A gapped Goldstone, $\theta$, with gap $\mu$:
\begin{align}
\omega^2_k=k^2+\mu^2\,. \label{dispi}
\end{align}
\item A radial mode, $\rho$, with gap $m_\rho^2=6\mu^2-2m^2$ and dispersion relation:
\begin{align}
\omega^2_k&=k^2+3\mu^2-m^2+\sqrt{(3\mu^2-m^2)^2+4k^2\mu^2} \notag \\
&=6\mu^2-2m^2+\frac{5\mu^2-m^2}{3\mu^2-m^2}k^2+\mO\left(\frac{k^4}{\mu^2}\right)\,. \notag
\end{align}
\end{itemize}
The masses of the gapless and the gapped Goldstone are fixed by symmetry and cannot be renormalized by loop effects \cite{Nicolis_Theorem,Brauner1loop}. Notice that due to the mixing term in \eqref{eqTripletLagrangianQuadratic}, the radial mode and the gapless one are interpolated both by $\psi$ and $h$---which decouple only for $\mu=0$. 

Finally note that if chemical potential is large enough, $\mu^2 \gtrsim |m^2|$, the mass of the radial mode and that of the gapped Goldstone can be of the same order, $m_\rho\sim\mu$. At low energies, $m_\rho$ sets the cutoff of the standard {\emph{quasi-relativistic}} effective theory for the Goldstone bosons. In the setup we are considering, the gapped Goldstones might hence lie outside the regime of validity of such EFT.

\subsection{Interactions of slow gapped Goldstones} \label{SecTripletProcesses}

Given the action~\eqref{eqTripletLagrangianQuadratic} we can now compute the amplitudes for the two processes involving the gapped Goldstone on the external legs: the $\theta\theta\to\theta\theta$ scattering and the $\theta\theta\to\pi_3\pi_3$ annihilation. We examine the amplitudes in the limit when the gapped Goldstones are slow. The reason for doing that is twofold. First, we verify the existence of Adler's zero in the amplitudes when one of the gapped Goldstones is at rest. Note that the interaction strength is not manifestly controlled by the gapped Goldstone's {3-momentum}. Consequently, {when the latter vanishes}, the amplitude does not vanish diagram by diagram, but only once all of them are taken into account. Second, we will use these results as our reference point to match the NREFT we will build in the next sections. In particular, the second process does not preserve the number of gapped Goldstones and, as anticipated in the Introduction, will be included in the NREFT through an imaginary part for some of the effective coefficients.

Note that, because of the kinetic mixing between $h$ and $\psi$, the calculation of the scattering amplitude is rather tedious (but straightforward). We spare the reader the details.

Consider first the elastic scattering, $\theta(\bm p_a)+\theta(\bm p_b)\rightarrow\theta(\bm p_c)+\theta(\bm p_d)$, in the limit {where} the gapped Goldstones are slow. {In the presence of a slow massive particle, it is customary to power-count interactions in terms of its velocity, $v\ll 1$ \cite{Luke}, which is related to its momentum and kinetic energy by $\bm p=\mu \bm v$ and $\epsilon=\omega-\mu\sim\mu v^2$.}
We {then} expand the tree-level matrix element for the scattering in powers of velocity:
\begin{align}
\mM=\mM^{(1)}+\mM^{(2)}+\ldots\,,\qquad \text{with} \qquad \mM^{(n)}\sim \mO(\bm p^{2n}/\mu^{2n})\,.
\end{align}
The leading order contribution is $\mO(v^2)$ and is given by
\begin{align}\label{eqTripletScattering1}
\mM^{(1)}=
\frac{\lambda}{\mu^2-m^2}\left[
\frac{(\bm{p}_a^{\,2}-\bm{p}_c^{\,2})^2}{(\bm{p}_a-\bm{p}_c)^2}+\frac{(\bm{p}_a^{\,2}-\bm{p}_d^{\,2})^2}{(\bm{p}_a-\bm{p}_d)^2}
-(\bm{p}_a+\bm{p}_b)^2\right]\,.
\end{align}
Setting one of the momenta to zero, say $\bm p_a=0$, this amplitude vanishes by conservation of energy, which implies $\bm p_b^2=\bm p_c^2+\bm p_d^2$  at the lowest order in velocity.
Notice also that the amplitude is {bounded albeit discontinuous}
in the collinear limits, $\bm p_a\rightarrow\bm p_c$ {or $\bm p_a\rightarrow\bm p_d$.}

For the purpose of matching with the NREFT it is also instructive to compute the next order amplitude, which reads
\begin{align}
\mM^{(2)}&=
\frac{\lambda}{\mu^2(\mu^2-m^2)}\Bigg\{
\frac{\mu^2}{\mu^2-m^2}
\left(\bm{p}_a^{\,2}\bm{p}_b^{\,2}+\bm{p}_c^{\,2}\bm{p}_d^{\,2}\right)
-
\frac{\mu^2+m^2}{4(\mu^2-m^2)}(\bm p_a^2+\bm p_b^2)^2
\notag \\&
\quad+\frac{7\mu^2+m^2}{\mu^2-m^2}(\bm{p}_a\cdot\bm{p}_b)^2
+\frac{2\mu^2}{\mu^2-m^2}\left[
(\bm{p}_a\cdot\bm{p}_c)(\bm{p}_b\cdot\bm{p}_d)
+(\bm{p}_a\cdot\bm{p}_d)(\bm{p}_b\cdot\bm{p}_c)\right]
\notag \\&
\quad-\frac{2\mu^2}{\mu^2-m^2}(\bm{p}_a^{\,2}+\bm{p}_b^{\,2}) (\bm{p}_a\cdot\bm{p}_b)
\label{eqTripletScattering2} \\&
\quad+\frac{(\bm{p}_a^{\,2}-\bm{p}_c^{\,2})^2}{(\bm{p}_a-\bm{p}_c)^2}
\left[\frac{3 \mu^2-m^2}{4(\mu^2-m^2)}\frac{(\bm{p}_a^{\,2}-\bm{p}_c^{\,2})^2}{(\bm{p}_a-\bm{p}_c)^2}-\frac{1}{2}(\bm p_a^2+\bm p_b^2)
+\frac{1}{2}\frac{\bm{p}_a^{\,2}\bm{p}_b^{\,2}-\bm{p}_c^{\,2}\bm{p}_d^{\,2}}{(\bm{p}_a^{\,2}-\bm{p}_c^{\,2})}\right]
\notag \\&
\quad+\frac{(\bm{p}_a^{\,2}-\bm{p}_d^{\,2})^2}{(\bm{p}_a-\bm{p}_d)^2}
\left[\frac{3 \mu^2-m^2}{4(\mu^2-m^2)}\frac{(\bm{p}_a^{\,2}-\bm{p}_d^{\,2})^2}{(\bm{p}_a-\bm{p}_d)^2}-\frac{1}{2}(\bm p^2_a+\bm p^2_b)
+\frac{1}{2}
\frac{\bm{p}_a^{\,2}\bm{p}_b^{\,2}-\bm{p}_c^{\,2}\bm{p}_d^{\,2}}{(\bm{p}_a^{\,2}-\bm{p}_d^{\,2})}\right]\Bigg\}\,. \notag
\end{align}
Again one can check the existence {of Adler's zero} when one of the 3-momenta vanishes. Note also that 
$s$-channel exchange of the radial mode $\rho$ gives terms whose expansion in 
{momenta} is controlled by $ \frac{\bm p ^2}{m_\rho^2-4 \mu^2}\propto \frac{\bm p ^2}{\mu^2-m^2}$. The expansion therefore breaks down {for momenta $p\sim\sqrt{\mu^2-m^2}$ or, alternatively, in the limit $\mu^2 \to m^2$ where the expectation value $\phi_0^2\propto(\mu^2-m^2)$ vanishes and the symmetry is restored.}

Since the internal symmetry group is fully broken, there is no symmetry left to protect the number of gapped Goldstones. 
Indeed, two of them may annihilate into two gapless Goldstones via the process $\theta(\bm p_a)+\theta(\bm p_b)\rightarrow \pi_3(\bm k_a)+\pi_3(\bm k_b)$. {Since the gapped Goldstones have energies $\geq \mu$, the final products of this annihilation have momenta and energies $\geq \mu$. Consequently,  in the regime $\mu\sim m_\rho$, this process is beyond the regime of applicability of an ordinary low-energy EFT.}

At the leading order in the gapped Goldstones' velocities the annihilation amplitude reads
\begin{align}\label{eqTripletAnnihilatioMatrixElement}
\mM=\frac{\lambda}{\mu^2-m^2}\left[\alpha \, (\bm p_a\cdot\bm p_b)
+ \beta \, \frac{(\bm p_a\cdot\bm k)(\bm p_b\cdot\bm k)}{\mu^2} \right]
 + \mO\left(\frac{\bm p^{2} (\bm p\cdot\bm k)}{\mu^4}\right),
\end{align}
where at the lowest order $\bm k_a= -\bm k_b\equiv \bm{k}$, with $|\bm k| = \mu$, and the dimensionless coefficients $\alpha$ and $\beta$ can be found in appendix~\ref{AppTripletAnnihilation}. Once again the amplitude vanishes when either initial 3-momenta is set to zero.
The leading order total annihilation cross section reads
\begin{align}\label{eqTripletAnnihilationCrossSection}
\sigma_\text{ann}\simeq\frac{1}{2\mu|\bm p_a-\bm p_b|}
\left[\left(\gamma+\delta\right)\frac{ (\bm p_a\cdot\bm p_b)^2}{\mu^4}+\delta \, 
\frac{\bm p^2_a\,\bm p^2_b}{\mu^4}\right],
\end{align}
where $\gamma$ and $\delta$ are dimensionless coefficients again given in appendix \ref{AppTripletAnnihilation}.

\subsubsection*{Intermezzo: gapped Goldstone decay}

{Notice that $\theta$ is odd under the unbroken $\mathds{Z}_2$ symmetry.  Processes with an odd number of  $\theta$ legs are thus forbidden, and $\theta$ is stable. The  $\mathds{Z}_2$ symmetry is an accident of the simple model under consideration and not a structural property of gapped Goldstones. 
That is appreciated, for instance, by showing that the addition of a new field allows to write  $\mathds{Z}_2$-breaking terms  and induce $\theta$-decay}---see appendix~\ref{AppTripletDecay} for an explicit construction using a complex $U(2)$ doublet. One finds that the decay amplitude vanishes when the 
{3-momentum} of $\theta$ approaches zero. The total decay rate for a gapped Goldstone with momentum $\bm p$ to leading order in velocity reads
\begin{align}\label{eqTripletDoubletDecay}
\Gamma=c\,\frac{\bm p^2}{\mu},
\end{align}
where $c$ is a dimensionless coefficient which depends on the couplings. 

\vspace{1em}

{In summary, just like for standard Goldstones, the interaction strength of gapped Goldstones is set by their spatial momentum.}
This is due to the fact that the zero mode of $\theta$ is not dynamical, but corresponds to a symmetry transformation of the vacuum, as discussed in section~\ref{sec:themodel}. More precisely, one can prove the existence of Adler's zeros~\cite{Weinberg2} for the matrix elements of gapped Goldstones at rest~\cite{Brauner}. Lastly, since no symmetry protects the number of gapped Goldstones, they may decay and/or annihilate into final states with energies of order $\mu$. When $\mu\sim m_\rho$ such final states cannot be described within any low-energy EFT, which is valid at energies much smaller than $m_\rho$ itself. { However, in this very situation,  the decay and annihilation processes happen within  a short distance scale. As  we shall see, that allows to consistently describe these effects via local operators in the  NREFT. These operators are  however  non-Hermitian, which makes the NREFT non-unitary.}

\section{The Nonrelativistic EFT: the universal description of slowly moving gapped Goldstones}\label{SecNREFT}

In the presence of spontaneous symmetry breaking one expects the low-energy dynamics to be effectively describable in terms of symmetries, and through a systematic derivative expansion. Such a construction (also known in jargon as coset or CCWZ construction) is expected to apply universally, i.e. purely on the basis of the symmetry breaking pattern and independently of the details of the underlying microscopic physics. In the known examples, it applies equally well to cases that purely involve the breaking of internal symmetries~\cite{CCWZ1,CCWZ2}, and to cases that involve the breaking of the spacetime ones (see, e.g.,~\cite{ogievetsky1974nonlinear,MoninWheel,Nicolis_More}). 

In the presence of gapped Goldstone bosons the situation can  however be more  involved. {That depends on the existence of two in principle distinguished scales:
the chemical potential $\mu$, which controls the gap of some of the Goldstones, and the scale $\Lambda$  which controls the gap of non-Goldstone degrees of freedom, as well as the derivative expansion.}\footnote{We are working under the simplifying assumption that the typical speed of the excitations around the cut of scale $\Lambda$ are $\mathcal{O}(1)$ so that there is no need to distinguish energy and momentum cutoffs.} The existence of a hierarchy, $\mu\ll \Lambda$, should generically correspond to the existence and smoothness of the limit 
 $\mu\to 0$, where  the charge density goes to zero, Lorentz invariance is recovered and the Goldstone bosons are the only light modes.
 An example of this situation is given by the linear $\sigma$-model of the previous section for the choice $m^2<0$, where the symmetry is broken already at $\mu=0$, where the density vanishes.
Generically, $\mu\ll\Lambda$ thus corresponds to the situation where the internal symmetry is partially broken already at zero density, and where  the state with finite charge density (and the corresponding Lorentz breaking) is fully described as a particular solution of  the original relativistic Goldstone EFT.
  Previous studies of the finite density systems based on the EFT methods~\cite{Nicolis_More,WatanabeMNGB,Brauner} have all focused on this case. In this setup the construction of the effective Lagrangian for the Goldstones proceeds in a way similar to the Lorentz invariant case, where there is a well defined derivative expansion, whose strength is controlled by $\Lambda$ itself. {For $\mu\ll \Lambda$, besides the counting of Goldstone degrees of freedom, there are no major structural novelties with respect to the standard relativistic CCWZ construction.}
  
The novelties appear when there is basically a single mass scale, $\mu\sim\Lambda$, {which is indeed} a minimal option for a system at finite density. Again, intuitively this regime corresponds to the situation where all symmetry breaking is fully dominated by the presence of finite
density. The limit $\mu\to 0$ cannot therefore be smooth. An example of this situation is given by the linear $\sigma$-model in the regime
$\mu^2\gg m^2>0$, where $\mu$ controls both the gap of the Goldstones and the gap of the radial non-Goldstone mode $\rho$. In fact, this situation is unavoidably realized whenever the system is (approximately) scale invariant with  $\mu$ representing the dominant  {\it spontaneous} source of breaking of scale invariance. This class of systems includes the physically relevant cases of conformal field theories (CFTs) in the large charge regime~\cite{Hellerman,MoninCFT,Bern1,HellermanO41},  and finite density QCD with large isospin chemical potential $\mu_I\gtrsim \Lambda_\text{QCD}$~\cite{Kaplan_KaonCondensate,Brown_NeutronStar,Son_Kaon1,Son_Kaon2}.  

The goal of this section is to present general,  systematic and self-consistent rules for constructing  the effective Lagrangian.
The relevant degrees of freedom will be the small 3-momentum modes: \emph{soft} gapless  and \emph{slow} gapped.
The first step will be to show explicitly how to organize the derivative expansion,  which involves of course both time and space derivatives, as  an expansion in the 3-momentum.
Secondly, we will have to properly interpret the result according to the rules of nonrelativistic EFTs. In particular, the conservation of the number of
 gapped Goldstones will emerge as a formal symmetry of the effective action. Processes where the gapped Goldstone number is not conserved will then  be described consistently, {but in an inclusive manner only}, by allowing for absorbitive imaginary coefficients in the effective Lagrangian.

We shall focus on the general class of models where  a global $SU(2)$ is nonlinearly realized at finite chemical potential $\mu$. The triplet model discussed in the previous section is a particular weakly coupled renormalizable example. {It will serve as  template and test case for our results.} Our discussion wants to be general, and applies in particular  to the case $\mu \sim \Lambda$. {In fact, our EFT construction will even 
apply to the case where non-Goldstone degrees of freedom with gap $\Lambda \ll \mu$  have been integrated out. However for economy of thought we shall mostly stick to the case $\mu \sim \Lambda$ when picturing our scenario.
Under our assumptions, any process where the number of gapped Goldstones is not conserved necessarily leads to the production
of states with momentum $\sim\mu$ (either gapless Goldstones or non-Goldstone states with gap less than $\mu$) that lie outside the domain of validity of the EFT. }  Our effective Lagrangian must thus necessarily be endowed with an effectively conserved gapped Goldstone number. We will concretely see how this happens.

As specified in the Introduction, we are interested in systems which spontaneously break an $SU(2)$ internal symmetry, as well as time translations and boosts, leaving unbroken the combination $\bar H = H + \mu Q_3$. 
In general we could parametrize the degrees of freedom of our EFT using the coset construction generalized to include spacetime symmetries~\cite{ogievetsky1974nonlinear,MoninWheel,Nicolis_More}.
This construction is illustrated in appendix~\ref{sec:cosetgeneral}.
We however find it more convenient to employ an equivalent approach: we define our fields in terms of the Lorentz-preserving {$SU(2)$} coset  
which involves three Goldstone fields, and then consider a generic time-dependent solution which further breaks spacetime symmetries down to spatial rotations, spatial translations and the modified time translation $\bar H = H + \mu Q_3$. 

Our dynamical variable just corresponds to a general $SU(2)$ matrix, $\Omega(x)$,
on which the group acts on the left:
\begin{align}
\Omega(x)\,\to \, g\Omega(x)\,,\qquad\qquad g\in SU(2)\,.
\end{align}
We can now choose local Lie parameters, the Goldstone fields, to parametrize $\Omega$. We will work with two different parametrizations, each showing advantages and disadvantages. The first parametrization, which we will name ``Left'', is
\begin{align} \label{eq:coset1}
\Omega(\chi,\alpha) = e^{i \chi Q_3} e^{i \alpha \frac{Q_+}{2} + i \alpha^* \frac{Q_-}{2}} \equiv e^{i \chi Q_3}\,  \Omega_L(\alpha)\,,
\end{align}
where $\chi$ and $\alpha\equiv\alpha_1+i\alpha_2$, represent the three real Goldstone scalars, and $Q_\pm\equiv Q_1\pm i Q_2$. Notice that $\Omega_L$ parametrizes the coset $SU(2)/U_3(1)$, with obvious notation. The other parametrization, which we dub as ``Right'', is instead
\begin{align} \label{eq:coset2}
\Omega(\chi,\pi) = e^{i \pi\frac{Q_+}{2} + i \pi^* \frac{Q_-}{2}}  e^{i \chi Q_3}\equiv \Omega_R(\pi) \, e^{i \chi Q_3}   \,,
\end{align}
with similar comments. The mapping between Left and Right parametrization is simply given by $\pi = e^{i\chi} \alpha$.

\subsection{Building the EFT with the Left parametrization} \label{SecNREFTCoset}

The CCWZ prescription \cite{CCWZ1,CCWZ2} allows to construct an $SU(2)$ invariant Lagrangian for the Goldstone fields $\chi$ and $\alpha$. Explicitly, the Maurer-Cartan one-form defines the covariant derivatives of the Goldstones \cite{MoninWheel} as
\begin{align}\notag
\Omega^{-1}\partial_\mu\Omega&=
i\partial_\mu\chi\, \Omega^{-1}_LQ_3\Omega_L+\Omega^{-1}_L\pd_\mu\Omega_L \\
&\equiv iD_\mu\chi Q_3+iD_\mu\alpha\frac{Q_+}{2}+iD_\mu\alpha^*\frac{Q_-}{2}\,,
\label{eq_cov_dv}
\end{align}
where
\begin{align}\label{eqCovariantDerivativePi3}
D_\mu\chi &=\pd_\mu\chi\cos\left(|\alpha|\right)+\frac{i\alpha^*\pd_\mu\alpha- i\alpha\pd_\mu\alpha^*}{|\alpha|^2}\sin^2\left(|\alpha|/2\right)
\,,
\\ \label{eqCovariantDerivativePi}
D_\mu\alpha 
&=i\pd_\mu\chi\,\alpha\frac{\sin\left( |\alpha|\right)}{|\alpha|}+
\frac{1}{2}\pd_\mu\alpha
\left(1+\frac{\sin|\alpha|}{|\alpha|}\right)+
\frac{\alpha}{2\alpha^*}\pd_\mu\alpha^*\left(
1-\frac{\sin|\alpha|}{|\alpha|}
\right)\,.
\end{align}

Then the most general $SU(2)$ invariant Lagrangian for $\chi$ and $\alpha$ is an arbitrary function of the covariant derivatives in \eqref{eq_cov_dv} and $\pd_\mu$:
\begin{align}\label{eq_GenLag}
\mathcal{L}=F\left[D_\mu\chi,D_\mu\alpha,D_\mu\alpha^*,\pd_\mu\right]\,,
\end{align}
with spacetime indices contracted in a Lorentz invariant way.

We are interested in a setup where spacetime symmetries are spontaneously broken as well. To this aim, we notice that the equations of motion deriving from \eqref{eq_GenLag} generically admit a solution of the form
\begin{align}\label{eq_GenSol0}
\chi=\mu t\,,\qquad\alpha=v\,,
\end{align}
where $v$ is a constant whose value depends on $\mu$. This is particularly easy to show using the Left parametrization \eqref{eq:coset1}.
Indeed, 
the Euler-Lagrange equation for the field $\chi$ takes the form
\begin{align}
-\pd_\mu\frac{\pd \mathcal{L}}{\pd (\pd_\mu\chi)}+
\pd_\mu\pd_\nu
\frac{\pd \mathcal{L}}{\pd (\pd_\mu\pd_\nu\chi)}+\ldots=0\,,
\end{align}
which is automatically satisfied since the Lagrangian and its derivatives do not depend on $x$ on the ansatz \eqref{eq_GenSol0}. Similarly, the only nontrivial contribution from the equation for $\alpha$ is
\begin{multline}
\frac{\pd\mL}{\pd\alpha}=\mu\left\{-\frac{v^*\sin(|v|)}{2|v|}\frac{\pd F}{\pd D_0\chi}
+\frac{1}{2}\left[\cos(|v|)+\frac{\sin (|v|)}{|v|}\right]
\frac{\pd F}{\pd D_0\alpha}\right.\\
\left.+
\frac{v^*}{2v}\left[\cos(|v|)-\frac{\sin (|v|)}{|v|}\right]
\frac{\pd F}{\pd D_0\alpha^*}\right\}=0\,,
\end{multline}
where the derivatives of the Lagrangian are evaluated on the ansatz. This is an algebraic equation determining the complex value of $v\equiv v(\mu)$.

It is convenient to perform a field redefinition of the form
\begin{align}\label{eqFieldRedef}
\Omega(\chi,\alpha)\equiv \Omega(\chi',\alpha')\exp\left[i v \frac{Q_+}{2} + i v^* \frac{Q_-}{2}\right]
\end{align}
to bring the solution \eqref{eq_GenSol0} to the form 
\begin{align}\label{eq_GenSol}
\chi'=\mu t\,,\qquad
\alpha'=0\,.
\end{align}
With the field redefinition \eqref{eqFieldRedef}, the covariant derivatives in \eqref{eq_cov_dv} are a linear combination of the ones for $\chi'$ and $\alpha'$, computed from $\Omega^{-1}(\chi',\alpha')\pd_\mu\Omega(\chi',\alpha')$.
Hence, by redefining its coefficients, the Lagrangian \eqref{eq_GenLag} takes an analogous form in terms of the fields $\chi'$ and $\alpha'$, and we can work equivalently with the primed fields. {The use of the primed variables corresponds to the request of tadpole cancellation imposed in ref.~\cite{Nicolis_More}.\footnote{In ref.~\cite{MoninCFT} it was wrongly concluded that tadpoles imply a deviation of the gap from $\mu$. This wrong conclusion has however no further consequence on the results there derived.} } In the following we shall drop the prime superscript.

The solution \eqref{eq_GenSol} spontaneously breaks time translations and boosts while being invariant under the action of $\bar{H}$. Therefore, to explicitly realize a symmetry breaking pattern of the desired form it is enough to expand the generic Lagrangian in \eqref{eq_GenLag} around the background \eqref{eq_GenSol}.

Notice that in this way of proceeding we did not need to introduce Goldstone fields for the broken boost generators. It is indeed known that, in order to realize spacetime symmetries nonlinearly, one {normally} needs fewer Goldstones than the number of broken generators \cite{LowIHC}. In the procedure detailed in appendix~\ref{sec:cosetgeneral}, where one introduces a coset parametrizing the full spacetime symmetry group~\cite{ogievetsky1974nonlinear,MoninWheel,Nicolis_More}, the boost Goldstone bosons are eliminated via an inverse Higgs constraint~\cite{IvanovIHC}. The final result is equivalent to the simple construction presented above.

The field parametrization in Eq. \eqref{eq:coset1}, expanded around the background \eqref{eq_GenSol}, makes clear the origin of the gap for the massive Goldstone.
Indeed, as a consequence of the $SU(2)$ symmetry, the Goldstone fields admit a solution where $\chi(x)=\mu t$ and the $\alpha$ field oscillate in time with frequency $\mu$. To see this, it is enough to show that such a configuration {is generated} by a symmetry transformation of the background \eqref{eq_GenSol}. Acting with a rotation generated by, say, $Q_1$ on the coset element one gets
{\begin{align}\label{eq_rot_gap}
\begin{split}
e^{i\xi Q_1} \Omega(\chi,\alpha) &= e^{i\chi Q_3} \left( e^{-i\chi Q_3} e^{i\xi Q_1} e^{i \chi Q_3} \right) \Omega_{L}(\alpha) \\
& = e^{i\chi Q_3} e^{i\xi\left(e^{-i\chi}\frac{Q_+}{2}+e^{i\chi}\frac{Q_-}{2}\right)} \Omega_{L}(\alpha) \equiv e^{i\tilde \chi Q_3} \Omega_{L}(\tilde \alpha)\,.
\end{split}
\end{align}
When one acts on the background $\chi=\mu t $ and $\alpha=0$, the transformed field, $\tilde \alpha=e^{-i\mu t}\xi$, is} {oscillating with frequency $\mu$.}

When spacetime symmetries are unbroken, the Goldstone fields transform with a constant shift under {an infinitesimal  group transformation} of the background. Standard relativistic EFTs describe the dynamics of slowly varying fields, corresponding to those configurations which are indistinguishable from a symmetry transformation at short distances.
The situation is quite different when considering a background of the form \eqref{eq_GenSol}. Indeed, we saw in Eq.~\eqref{eq_rot_gap} that an $SU(2)$ rotation can generate a configuration oscillating in time with a frequency of the order of the cutoff of the theory. {This is the main disadvantage of the Left parametrization. Then, to proceed formulating the EFT, it is more convenient to use the alternative field parametrization~\eqref{eq:coset2}, for which the group action takes a different form.}


\vspace{1em}
\subsection{Building the EFT with the Right parametrization} \label{SecNREFTCosetRight}

In the field parametrization~\eqref{eq:coset2}, the background solution reads as in~\eqref{eq_GenSol}:
\begin{align}
\chi=\mu t+\pi_3\,,\qquad
\pi_3=\pi=0\,.
\end{align}
However, the group action takes now a different form.  {As a result, a generic {infinitesimal} $SU(2)$ transformation acting on the background provides a solution of the form $\pi_3=\text{constant}$ and $\pi=\text{constant}$, precisely like in a Poincar\`e invariant coset. In analogy with that case the EFT
will thus be limited to the slowly varying field configurations,  $\pd\pi\ll\mu\pi\,,\;\pd\pi_3\ll\mu \pi_3$, in the Right parametrization \eqref{eq:coset2}.}

Notice that, despite $\pi=\text{constant}$ {being} a solution, the field $\pi$ describes a gapped mode with frequency $\mu$. To see this, recall that the gap is measured by the action of the unbroken generator of time translations: $\bar{H}=H+\mu Q_3$. It is then possible to verify that under the action of $\bar{H}$, the field acquires a phase proportional to $\mu$: $\pi(t,{\bm x})\rightarrow e^{-i\mu\delta t }\pi(t+\delta t,{\bm x})$. Thus, in this parametrization, low frequency modes for the field $\pi$ are associated with \emph{slowly moving} gapped Goldstones.
{The EFT  thus consists of modes with small 3-momentum, and with eigenvalues of $\bar{H}=H+\mu Q_3$ around respectively $0$ for $\pi_3$ and $\mu$ for $\pi$. Modes that do not satisfy these requirements should be thought as having been integrated out.}

{Because of the {unusual transformation property of the field $\pi$ under the unbroken time translations, the Lagrangian \eqref{eq_GenLag},
written in the Right parametrization, is correspondingly unusual}: it is explicitly time dependent when expanded in fluctuations around \eqref{eq_GenSol}. To see this explicitly, let us compute the Maurer-Cartan one-form.} 
Using \eqref{eq:coset2}, we write it as follows
\begin{align} \label{eq:timedep}
\Omega^{-1}\partial_\mu\Omega &= e^{-i\chi Q_3} \Omega_{R}^{-1}\partial_\mu \Omega_{R} e^{i\chi Q_3} + i\partial_\mu\chi Q_3 \notag \\
& = e^{-i\chi Q_3} \left( i d_\mu \pi \frac{Q_+}{2} + i d_\mu \pi^* \frac{Q_-}{2} + i A_\mu Q_3 \right) e^{i\chi Q_3} + i\partial_\mu\chi Q_3 \notag \\
& = i\left( e^{-i\chi} d_\mu\pi \frac{Q_+}{2} + e^{i\chi} d_\mu\pi^* \frac{Q_-}{2} + D_\mu\chi Q_3 \right)\,.
\end{align}
Here $d_\mu\pi$ and $A_\mu$ are the covariant derivative and the connection for the $SU(2)/U_3(1)$ coset, given by
\begin{align}
d_\mu\pi &= \pi\frac{\pi^*\partial_\mu\pi-\pi\partial_\mu \pi^*}{2{|\pi|}^3}\sin\big(|\pi|\big) +\pi\frac{\pi^*\partial_\mu\pi+\pi\partial_\mu\pi^*}{2{|\pi|}^2}\,,\\
A_\mu &= i\frac{\pi^*\partial_\mu\pi-\pi\partial_\mu\pi^*}{{|\pi|}^2}\sin^2\big(|\pi|/2\big)\,.
\end{align} 
The full $SU(2)$ covariant derivatives \eqref{eq_cov_dv} are written in terms of these as
\begin{align}\label{eq:CD}
D_\mu\alpha=e^{-i\chi} d_\mu\pi\,,\qquad
D_\mu\chi = \partial_\mu\chi + A_\mu \,.
\end{align}
{By Eqs.~\eqref{eq:timedep}-\eqref{eq:CD} a generic invariant Lagrangian, through the factor $e^{i\chi}$, contains terms that  explicitly depend on time on the background. This seems a rather unpleasant property. However one must keep in mind that our EFT  only contains  low frequency/low momentum   modes ($\pd\pi\ll\mu\pi\,,\;\pd\pi_3\ll\mu \pi_3$). 
Then, by simple Fourier analysis,   Lagrangian terms involving 
a non-trivial  power of  $e^{i\chi}$   integrate to zero in the action, as its fast oscillation cannot be compensated by any finite combination of EFT modes.
%
Only terms featuring no power of $e^{i\chi}$ survive. These are   invariant under an emergent $U(1)$ {symmetry, $U_\pi(1)$, } acting as $d_\mu\pi\to e^{i\xi}d_\mu\pi$,\footnote{This coincides with the $U(1)$ generated by the action of $Q_3$ on the right of the coset.} which is nothing but the particle number conservation of nonrelativistic theories} (see e.g.~\cite{GuthNRLimit}). {As typical of a nonrelativistic limit, this property emerges naturally after factoring out the mass contribution from the time evolution of the gapped fields, as we did switching from the Left to the Right parametrization.}

The emergence of this $U(1)$ symmetry does not allow to describe processes where the number of gapped Goldstones is changed, such as decay or annihilation. Physically this is because they necessarily feature modes with {momentum} $\sim \mu$ in the final state, outside the regime of validity of the effective theory. As a consequence the resulting nonrelativistic EFT cannot be unitary. {Indeed, through  the optical theorem, these processes give rise to imaginary parts in the gapped Goldstone propagators and matrix elements, which can only be matched in the nonrelativistic EFT by allowing for imaginary parts in the Wilson coefficients \cite{Braaten}.} We will discuss this matching in some detail for the linear triplet model in the following sections.

\vspace{1em}

We would now like to expand the Lagrangian \eqref{eq_GenLag} in a series of higher derivative terms. {In order to power count, it is useful to indicate by $\partial_{s}\ll \mu$ the {\it small} derivatives of our EFT modes. More precisely, the spacial part $\bm{\partial}$ obviously represents the small momentum for both $ \pi_3$ and $\pi$, while $\partial_t$, represents respectively energy and kinetic energy for $\pi_3$ and $\pi$. Remember indeed that in the Right parametrization we have in practice subtracted $\mu$ from the oscillation frequency of $\pi$ excitations.}

 {The parametrization \eqref{eq:coset2} shows that the na\"ive derivative expansion must be reorganized when working around the typical background we are interested in.}
Consider, in fact, the derivative of the Maurer-Cartan form:
\begin{align} \label{eq:dMC}
\begin{split}
\partial_\mu \left[ \Omega^{-1} \partial_\nu \Omega \right] &= -i\partial_\mu\chi \left[ Q_3, \Omega^{-1}\partial_\nu \Omega \right] + e^{-i\chi Q_3} \partial_\mu \left(\Omega^{-1}_{R}\partial_\nu \Omega_{R} \right) e^{i\chi Q_3} \\
& \quad + i \partial_\mu\partial_\nu \chi Q_3\,.
\end{split}
\end{align}
{The last two terms  are genuinely  suppressed by two EFT derivatives, $\mathcal{O}(\partial_s^2)$. However, around the background $\chi=\mu t$,  the first term counts as a one-derivative term, $\mathcal{O}(\mu\partial_s)$, unsuppressed with respect to $\mu \, \Omega^{-1} \partial_\nu \Omega$. This shows 
that some reorganization of terms is needed in order to write the Lagrangian in a manifest expansion in powers of $\partial_s$. Notice for that purpose that the first term in \eqref{eq:dMC} is not a new independent object; instead, it is proportional to the commutator of $Q_3$ with the Maurer Cartan form \eqref{eq_cov_dv}. This indicates how to proceed:
one can simply subtract the first term on the right hand side of Eq.~\eqref{eq:dMC}, so that the remaining terms are $\mathcal{O}(\partial_s^2)$.} Although this term is not $SU(2)$ invariant, there is a simple $SU(2)$ invariant Lorentz vector that is proportional to $\partial_\mu \chi$ at linear order, i.e. $D_\mu\chi$. We therefore can define a \emph{nonrelativistic derivative} in the following way:
\begin{align}  \label{nonrelativisticD}
\hat \partial_\mu \equiv \partial_\mu+iD_\mu\chi\big[Q_3,\cdot\big]\,,
\end{align}
where by $\big[Q_3,\cdot\big]$ we mean the action of the commutator and the derivative is meant to act on the Maurer-Cartan form.\footnote{Formally, 
Eq.~\eqref{nonrelativisticD} corresponds to the covariant derivative for an $SU(2)$ gauge group acting \emph{on the right} of the coset~\eqref{eq:coset2}, with a gauge connection given by $A_\mu^I=\delta^I_3 D_\mu\chi$.}
{By its definition, the action of  any power of $\hat \partial$ on the Maurer-Cartan form  is suppressed by the corresponding power of $\partial_s$:
}
\begin{align}\label{nonrelativisticDprop}
\hd_{\mu_1}\cdots\hd_{\mu_n}
\left[\Omega^{-1}\pd_\nu\Omega\right]
\ll \mu\, \hd_{\mu_1}\cdots\hd_{\mu_{n-1}}\left[\Omega^{-1}\pd_\nu\Omega\right]\,.
\end{align}
{The action on the covariant derivatives of \eqref{eq:CD} reads}:
\begin{align}
\hd_\mu D_\nu\chi=\pd_\mu D_\nu\chi\;,\qquad
\hd_\mu D_\nu\alpha=(\pd_\mu+i D_\mu\chi)D_\nu\alpha=
e^{-i\chi}\left(\pd_\mu+i A_\mu\right)d_\nu\pi.
\end{align}

{Since the second term in Eq.~\eqref{nonrelativisticD} is not a new object, formulating} {the EFT in terms of $\hd$ just amounts to rearranging the terms in the action
so as to make the expansion in powers of $\partial_s$ manifest.}  The new derivative allows us to define a consistent power counting in {the small spatial momentum for both the gapless and gapped Goldstones.}

We remark that Eq. \eqref{nonrelativisticD} is not the only possible choice for the definition of the \emph{nonrelativistic} derivative. For instance, it is possible to multiply $D_\mu\chi$ by an arbitrary function of $\sqrt{D_\mu\chi D^\mu\chi}/\mu $ without affecting the property \eqref{nonrelativisticDprop}.

\vspace*{1em}

In summary, to construct an effective action for the Goldstones that is invariant under the full symmetry group $SU(2)\times\text{Poincar\`e}$, and that has a consistent expansion in the limit of slow gapped Goldstones one needs to $(i)$ use the coset construction to build terms that are manifestly invariant under the unbroken group, $(ii)$ consider only operators that are invariant under an additional $U_\pi(1)$ {particle conservation symmetry}, and $(iii)$ construct higher derivative terms using the nonrelativistic covariant derivative~\eqref{nonrelativisticD}. This recipe can be generalized to different symmetry breaking patterns.

 {At the lowest derivative order, one finds three invariants under Lorentz and $U_\pi(1)$: $D_\mu \chi D^\mu\chi$, $\big| D_\mu \chi D^\mu \alpha \big|^2$ and $D_\mu \alpha D^\mu \alpha^*$.} It is convenient to organize them in terms of operators whose expectation value vanishes on the background \eqref{eq_GenSol}. To match to the spacetime coset construction reported in appendix~\ref{sec:cosetgeneral}, we reorganize them in the following way:
\begin{align}\label{eq_invariants}
\begin{split}
\nabla_0\pi_3 \equiv \sqrt{D_\mu\chi D^\mu\chi} - \mu\,,  \qquad \big| &\nabla_0\alpha \big|^2 \equiv 
\frac{\big| D_\mu\chi \,D^\mu \alpha \big|^2}{D_\nu\chi D^\nu\chi}
=\frac{\big| D_\mu\chi \,d^\mu \pi \big|^2}{D_\nu\chi D^\nu\chi}\,, \\
\big| \nabla_i\alpha \big|^2 \equiv 
\frac{\big| D_\mu\chi \,D^\mu \alpha \big|^2}{D_\nu\chi D^\nu\chi} & - D_\mu \alpha \,D^\mu \alpha^*
=\frac{\big| D_\mu\chi \,d^\mu \pi \big|^2}{D_\nu\chi D^\nu\chi}  - d_\mu \pi \,d^\mu \pi^*\,.
\end{split}
\end{align}
At the leading order in derivatives, the effective nonrelativistic Lagrangian then takes the form:
\begin{align} \label{eqNREFTLagrangiand2}
\mL_\text{eff}=c^{(1)}\mu^3\nabla_0\pi_3,+c^{(2)}_1\mu^2(\nabla_0\pi_3)^2
+c^{(2)}_2\mu^2|\nabla_0\alpha|^2-c^{(2)}_3\mu^2|\nabla_i\alpha|^2
+\mO\left(\mu\hat{\pd}^3\right)\,.
\end{align}
The action up to the fourth order in derivatives is given in appendix~\ref{AppNREFTAction}. In the next sections we discuss the degrees of freedom in this EFT and illustrate the power counting by calculating several sample processes.

\subsection{The NREFT to quadratic order}\label{SecNREFTLag}

Let us expand the Lagrangian \eqref{eqNREFTLagrangiand2} to quadratic order in the fields:
\begin{align} \label{eqNREFTQuadraticL}
\begin{split}
\mL_\text{eff}
\supset&
c^{(2)}_1 \mu^2 (\pd_0\pi_3)^2 -
\frac1 2 c^{(1)} \mu^2 (\bm{\nabla}\pi_3)^2
+\frac1 4 c^{(1)}\mu^3 \left[i\pi^* \pd_0\pi+\text{c.c.}
\right] \\
&-c_3^{(2)}\mu^2|\bm{\nabla}\pi|^2+
c_2^{(2)} \mu^2 \left|\pd_0\pi\right|^2.
\end{split}
\end{align}
We focus on configurations with small derivatives.
From Eq.~\eqref{eqNREFTQuadraticL} one finds that $\pi_3$ interpolates a gapless mode with dispersion relation
\begin{align}\label{eqDispersionNGB}
\omega^2_{k}=c_s^2 \bm k^2+\mO\left(\bm k^{\,4}/\mu^2\right)\,,\qquad
c_s^2\equiv\frac{c^{(1)}}{2c^{(2)}_1 }\,.
\end{align}
The quantization of $\pi_3$ then proceeds as usual, i.e.
\begin{align}
\pi_3(x)=\frac{c_s}{\mu\sqrt{c^{(1)}}}\int \frac{d^3k}{(2\pi)^3\sqrt{2\omega_{k}}}
a_{\bm{k}} e^{-i\omega_k t+i\bm k\cdot\bm{x}}+\text{h.c.}\,,\quad
[a_{\bm{k}},a^\dagger_{\bm{p}}]=(2\pi)^3\delta^3(\bm k-\bm p)\,.
\end{align}
To quantize the $\pi$ field, we notice that the last term in~\eqref{eqNREFTQuadraticL} contains two time derivatives and can be treated as a higher derivative {perturbation of} the third one, which contains only one. Indeed, $\pi$ has the kinetic term of a nonrelativistic field and is quantized as
\begin{align}\label{eqQuantizationGappedNGB}
\pi(x)=\sqrt{\frac{2}{c^{(1)}\mu^3}}\int \frac{d^3 p}{(2\pi)^3}b_{\bm{p}} e^{-i\epsilon_{p} t
+i\bm p\cdot\bm{x}}\,,\qquad
[b_{\bm{p}},b^\dagger_{\bm{k}}]=(2\pi)^3\delta^3(\bm p-\bm k)\,,
\end{align} 
with dispersion relation given by
\begin{align}\label{eqDispersionGappedNGB}
\epsilon_{p}=c_m\frac{\bm p^2}{2\mu}+\mO\left(\bm p^{\,4}/\mu^3\right)\,,\qquad
c_m\equiv\frac{4 c_3^{(2)}}{ c^{(1)}}\,.
\end{align} 
As commented before, due to its transformation properties under $\bar{H}$, $\pi$ really is a gapped field. The ladder operator $b_{\bm{p}}^\dagger$  then creates a gapped Goldstone state with energy $E_p=\mu+\epsilon_p$.\footnote{For the sake of the discussion, we are momentarily considering a theory in which the gapped Goldstone cannot decay. }

The nonrelativistic complex field $\pi$ only contains annihilation operators (and $\pi^*$ contains only creation ones) and thus propagates one degree of freedom.\footnote{
{Alternatively one could use the equations of motion to eliminate one of the two real components of the field $\alpha=\alpha_1+i\alpha_2$ of the Right parametrization \eqref{eq:coset1} in terms of the other. Doing so would change the description of the gapped Goldstone mode from a complex field with one time derivative kinetic term to a two derivatives real scalar field. To leading order in derivatives, this procedure \emph{formally} coincides with imposing an extra inverse Higgs constraint of the form $\text{Re}[\nabla_0\alpha]=0$. The same inverse Higgs constraint, but with a different physical interpretation, was discussed in \cite{Nicolis_More} for the case in which the EFT cutoff is much larger than the chemical potential, $\Lambda\gg\mu$. We provide a more detailed discussion in appendix \ref{sec_IHC}. }
} As anticipated, the present effective theory describes a gapless mode and a nonrelativistic gapped mode. As a consistency check, one can see that including higher derivative corrections, such as the last term in~\eqref{eqNREFTQuadraticL} or terms constructed with~\eqref{nonrelativisticD}, generates both new poles as well as correction to the dispersion relation~\eqref{eqDispersionGappedNGB}. The new poles generically appear for frequency or momenta of order $\mu$ and are outside the regime of validity of our EFT; they should therefore be discarded. The corrections to the dispersion relation are instead higher order in the low-momentum expansion, showing that these additional terms can consistently be considered as perturbations {in the  EFT}.

\subsection{Gapped Goldstone number conservation and non-unitarity}

The NREFT enjoys a $U_\pi(1)$ invariance, $\pi \to e^{i\xi}\pi$, corresponding to particle number conservation for the gapped Goldstones. {As already remarked, this does not correspond to a symmetry of the microscopic theory, but it is rather a consequence of the small momentum  and energy
window which characterizes the degrees of freedom of our EFT. In particular the EFT does not contain degrees of freedom with energy and momentum such that the $\pi$ can decay or annihilate into them \cite{GuthNRLimit}. Hence the conservation of $\pi$-number. On the other hand, in the full theory these processes
will in general exist, with final states involving $\pi_3$ modes with momentum $\sim \mu$, and also, possibly, other non-Goldstone degrees of freedom with 
gap $\sim \mu$.

The EFT cannot describe the $\pi$ decay or annihilation processes \emph{exclusively}, since the final states have short wavelengths. It can however describe them \emph{inclusively}. Indeed, by the optical theorem,  these processes give rise to  imaginary parts in the $\pi$ propagator and matrix elements, which can be matched in the NREFT by assigning proper imaginary parts to the Wilson coefficients.} For instance, an imaginary part for the ``kinetic energy coefficient" $c_m$ corresponds to a decay width of the gapped Goldstone:
\begin{align}
\Gamma_p = - 2 \, \text{Im}\left[E_p\right]=
-\text{Im}\left[c_m\right]\frac{\bm p^2}{\mu}\,.
\end{align}
{Notice that the above momentum dependence}  matches the explicit result we found in Eq.~\eqref{eqTripletDoubletDecay}. The resulting theory is therefore non-unitary and is sometimes called a complex NREFT~\cite{Braaten}.

Physically, annihilation and decay can be matched by means of local terms since these processes are determined by short distance dynamics.
More precisely, to match the imaginary parts of the propagator or scattering amplitudes for a slow $\pi$ of the full theory via local terms in the NREFT, requires the latter to be analytical in the 
{spatial momentum}. This is expected to be true  as long as the relevant kinematic region is separated by a finite gap from {any excitation which was not} included in the NREFT.

Notice also that since the zero gapped Goldstone sector, $\pi=0$, of the theory reduces to an EFT of a single gapless superfluid Goldstone, which should be unitary, the effective coefficients that multiply operators which do not contain $D_\mu\pi$ should always be real. {Consistently, we will see that this is the case when we will match our EFT to the linear triplet in the next section.}

\subsection{Interactions and power counting} \label{SecNREFTtreelevel}

In this section we describe some interaction processes arising in the NREFT we built. {In particular, we focus on two peculiar aspects: power counting 
and non-unitarity.} The techniques described here are heavily inspired by nonrelativistic QED (NRQED)~\cite{NRQED} and nonrelativistic QCD (NRQCD)~\cite{NRQCD}, which describe the interactions of heavy fermions in the presence of light gauge fields. {Like in those theories, we will find convenient to power count amplitudes in powers of the velocity $\bm{v}\sim\bm{p}/\mu$ of the heavy field.}

Consider first the expansion of the covariant derivatives \eqref{eq:CD},
\begin{align}
\begin{split}
D_\mu\pi_3 &=\pd_\mu\pi_3+\frac{i\pi^*\partial_\mu\pi-i\pi\partial_\mu\pi^*}{4}-|\pi|^2\frac{i\pi^*\partial_\mu\pi-i\pi\partial_\mu\pi^*}{48}+\mO(\pi^6)\,, \\
D_\mu\pi  &=e^{-i\chi}\left[\pd_\mu\pi+i\pi\frac{i\pi^*\partial_\mu\pi-i\pi\partial_\mu\pi^*}{12}+\mO(\pi^5)\right]\,.
\end{split}
\end{align} 
We see that { all terms in the action display derivatives acting on all the  fields,
making manifest  the vanishing of the interaction strength  with the 3-momentum, or equivalently with the gapped Goldstone velocity, in agreement with the results in section~\ref{SecTripletProcesses}.}

In deriving the dispersion relation \eqref{eqDispersionGappedNGB}, we realized that time and space derivatives of the on-shell gapped Goldstone field scale differently---namely $\bm \nabla \pi\sim\mu v$ and $\partial_0\pi\sim\mu v^2$---and some care is thus required in power counting.\footnote{For processes involving only the gapless mode the power counting is similar to the relativistic case.} 
Indeed, even after subtracting the mass contribution, a simple power counting in derivatives $\partial/\mu$ does not distinguish between $v$ and $v^2$, retaining more terms than needed at a fixed order in $v$. As in NRQED and NRQCD, the power counting in velocity is complicated by the presence of states with two different forms of dispersion relation~\cite{HoangNRQCD,RothTasi}.

\vspace{1em}

{We will now match the results of our {NREFT to those} of the model presented in section~\ref{SecMassiveNGBreview}. In particular, this means  the gapped Goldstone is stable  and its dispersion relation real, which allows us to put its external legs on-shell.}

To facilitate power counting it is convenient to split each field in components with support on different regions {of phase space}~\cite{Luke,RothTasi,Gri1,Gri2}. In particular, we write $\pi_3=\pi_3^\text{s}+\pi_3^\text{p}+\pi_3^\text{us}$ and $\pi=\pi^\text{s}+\pi^\text{p}+\pi^\text{us}$, where the labels stand respectively for soft, potential and ultrasoft.
We \emph{define} the different components according to the scaling of their energy and momemtum (i.e. time and space derivatives) with velocity:\footnote{{Note that for off-shell Goldstones there is a fourth possibility, namely $(\omega,\bm k)\sim (\mu v,\mu v^2)$; this  never appears in scattering processes \cite{RothTasi}, but might be relevant in other contexts. For example, when an external probe coupled to the  system releases} finite energy but almost vanishing spatial momentum~\cite{Caputo}.}
\begin{align}\label{eqSoftUltraSoftPotential}
\begin{split}
\text{soft:} & \quad (\omega,\bm k)\sim (\mu v,\mu v)\,, \\
\text{potential:} & \quad (\omega, \bm k)\sim(\mu v^2,\mu v)\,, \\
\text{ultrasoft:} & \quad (\omega,\bm k)\sim (\mu v^2,\mu v^2)\,.
\end{split}
\end{align}
Note that on-shell gapless Goldstones are contained in both $\pi_3^\text{s}$ and $\pi_3^\text{us}$, while on-shell gapped Goldstones are contained in $\pi^\text{p}$. The potential mode for $\pi_3$, as well as the ultrasoft and soft modes of $\pi$, which are well within our EFT, are instead never on-shell. Indeed, they can be considered as auxiliary fields that can, in principle,  be integrated out. This is customarily done in  commonly studied non-relativistic EFTs, such as NRQCD, at the price of introducing non-local interaction vertices between the on-shell modes \cite{RothTasi}. However, in our case, since the nonlinear action of the internal $SU(2)$ group mixes the different modes of the Goldstone fields, we prefer to keep track of all the modes in the EFT, both the on-shell and the off-shell ones, and work with a local Lagrangian.\footnote{The situation bears some similarity with the case of supersymmetry.  For an action that includes the auxiliary fields, supersymmetry is manifest and the field transformations are independent of the Lagrangian. Upon integrating them  out, supersymmetry is preserved, but the field transformations  depend on the Lagrangian itself. Similarly, integrating out the off-shell modes from the Lagrangian~\eqref{eqNREFTLagrangiand2} makes the $SU(2)$ transformations of the remaining fields coupling-dependent.}

Our classification \eqref{eqSoftUltraSoftPotential} thus differs from the standard NRQCD language, where the off-shell modes for the massive field never appear. Moreover to 
simplify the notation we have  classified modes as soft, potential and ultrasoft purely according to the scaling of their energy and momentum, rather then by the scaling of their propagator. The result is that the propagators of $\pi_3^s,\pi_3^p,\pi_3^{us}$ scale respectively like $1/v^2,1/v^2,1/v^4$ while those of  $\pi^s,\pi^p,\pi^{us}$  respectively scale like 
$1/v,1/v^2,1/v^2$.



\vspace{1em}

The rules are now the following: for each process under consideration one has to determine which field is participating in the different  vertices of the diagrams, perform the expansion of $\pi$ and $\pi_3$ mentioned above, and determine what are the relevant interaction terms at the given order in velocity.  As already mentioned, the propagators of gapless and gapped Goldstones will feature different scalings in $v$ and thus the resulting power counting will differ.

For leading order applications, it might still be useful in practice to first extract Feynman rules in a $\pd/\mu$ expansion and perform $v$ counting only afterwards. In appendix \ref{AppNREFTFeynamRules} we provide a list of Feynman rules to leading order in $\pd/\mu$.


\vspace{1em}

\begin{figure}[t]
\centering
\includegraphics[scale=1.05]{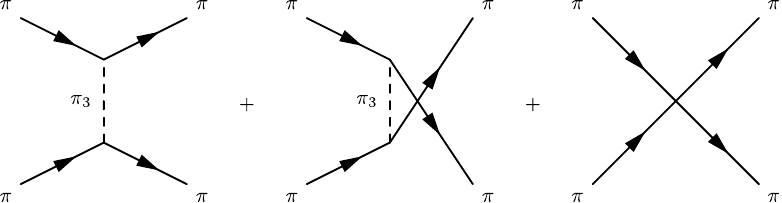}
\caption{Diagrams contributing to $\pi\pi\rightarrow\pi\pi$ at tree-level.}\label{FigPiPiscattering}
\end{figure}

Let us start discussing the $\pi(\bm{p}_a)+\pi(\bm{p}_b)\rightarrow \pi(\bm{p}_c)+\pi(\bm{p}_d)$ scattering at tree-level. In the NREFT only contact interactions and $\pi_3$ exchange diagrams contribute to this process, as in figure~\ref{FigPiPiscattering}.
By momentum conservation, the exchanged $\pi_3$ is an off-shell potential field.
{Given that,  the leading $\mO(v^2)$ amplitude is fully determined by the vertices of the leading Lagrangian ~\eqref{eqNREFTLagrangiand2} in the derivative expansion:}
\begin{align}
\begin{split}
\mL_\text{eff}&\supset
-\frac{i c_s}{2\mu^2\sqrt{c^{(1)}}}\big(\pi^{\text{p}\,*}\bm \nabla\pi^\text{p}-\pi^\text{p}\bm\nabla\pi^{\text{p}\,*}\big)\cdot\bm\nabla\pi_3^\text{p} \\
&\quad +\frac{(2 c_m-3)}{24 c^{(1)} \mu ^4}\big(\pi^{\text{p}\,*}\bm \nabla\pi^\text{p}-\pi^\text{p}\bm\nabla\pi^{\text{p}\,*}\big)^2
-\frac{|\pi^\text{p}|^2}{12 c^{(1)} \mu ^3}\big(\pi^{\text{p}\,*}\dot\pi^\text{p}-\pi^\text{p}\dot\pi^{\text{p}\,*}\big)\,,
\end{split}
\end{align}
where we canonically normalized fields as $\pi_3\rightarrow\frac{c_s}{\mu\sqrt{c^{(1)}}}\pi_3$ and $\pi\rightarrow\sqrt{\frac{2}{c^{(1)}\mu^3}}\pi$.
To order $\mO(v^2)$ the corresponding matrix element reads
\begin{align}\label{eqNREFTscattering1}
\begin{split}
\mM^{(1)}_\text{NR}&=\frac{1}{4c^{(1)}\mu^4}\bigg[
\frac{(\bm{p}_a^{\,2}-\bm{p}_c^{\,2})^2}{(\bm{p}_a-\bm{p}_c)^2}+\frac{(\bm{p}_a^{\,2}-\bm{p}_d^{\,2})^2}{(\bm{p}_a-\bm{p}_d)^2}
+\left(2c_m-3\right)(\bm{p}_a+\bm{p}_b)^2 \\
&\quad+2(1-c_m) \left(\bm{p}_a^{\,2}+\bm{p}_b^{\,2}\right)\bigg].
\end{split}
\end{align}
Once the coefficient $c_m$ is fixed by the dispersion relation~\eqref{eqDispersionGappedNGB}, this only depends  on the overall coefficient $c^{(1)}$. {Below we will match its value to the linear triplet model.
} Eq.~\eqref{eqNREFTscattering1} correctly vanishes in the limit where any of the gapped Goldstones is at rest, again in agreement with~\cite{Brauner}. One can similarly compute the $\mO(v^4)$ correction. To this end one has to consider the action up to the fourth order in covariant derivatives, which is presented in appendix~\ref{AppNREFTAction}. The resulting correction to the amplitude reads:
\begin{align}\label{eqNREFTscattering2}
\mM^{(2)}_\text{NR}&=\frac{1}{\mu^6[c^{(1)}]^2}\Bigg\{
\left(b_1-\frac{c^{(1)}c_m^2}{16c_s^2}\right)(\bm p^2_a+\bm p^2_b)^2
+\frac{c^{(1)}}{8}\left(\frac{c_m^2}{c_s^2}-c_m^{(2)}\right)
\left(\bm{p}_a^{\,2}\bm{p}_b^{\,2}+\bm{p}_c^{\,2}\bm{p}_d^{\,2}\right) \notag
\\& \quad
+b_2(\bm{p}_a^{\,2}+\bm{p}_b^{\,2})\bm{p}_a\cdot\bm{p}_b
+b_3(\bm{p}_a\cdot\bm{p}_b)^2 \notag 
\\& \quad
+b_4\left[(\bm{p}_a\cdot\bm{p}_c)(\bm{p}_b\cdot\bm{p}_d)
+(\bm{p}_a\cdot\bm{p}_d)(\bm{p}_b\cdot\bm{p}_c)\right] 
\\& \quad
+\frac{(\bm{p}_a^{\,2}-\bm{p}_c^{\,2})^2}{(\bm{p}_a-\bm{p}_c)^2}
\left[\frac{c^{(1)}c_m^2}{16c_s^2}\frac{(\bm{p}_a^{\,2}-\bm{p}_c^{\,2})^2}{(\bm{p}_a-\bm{p}_c)^2}-b_1(\bm p^2_a+\bm p^2_b)
+\frac{c^{(1)}c_m^{(2)}}{8c_m}
\frac{\bm{p}_a^{\,2}\bm{p}_b^{\,2}-\bm{p}_c^{\,2}\bm{p}_d^{\,2}}{(\bm{p}_a^{\,2}-\bm{p}_c^{\,2})}\right] \notag
\\& \quad
+\frac{(\bm{p}_a^{\,2}-\bm{p}_d^{\,2})^2}{(\bm{p}_a-\bm{p}_d)^2}
\left[\frac{c^{(1)}c_m^2}{16c_s^2}\frac{(\bm{p}_a^{\,2}-\bm{p}_d^{\,2})^2}{(\bm{p}_a-\bm{p}_d)^2}-b_1(\bm p^2_a+\bm p^2_b)
+\frac{c^{(1)}c_m^{(2)}}{8c_m}
\frac{\bm{p}_a^{\,2}\bm{p}_b^{\,2}-\bm{p}_c^{\,2}\bm{p}_d^{\,2}}{(\bm{p}_a^{\,2}-\bm{p}_d^{\,2})}\right]\Bigg\}\,. \notag
\end{align}
Here $c_s^2$ is defined in \eqref{eqDispersionNGB} and $c_m^{(2)}$ is {defined by  the gapped Goldstone dispersion relation at subleading order}~\eqref{eqDispersionGappedNGB}
\begin{align}\label{eqNREFTdispersionGappedNGB2}
\epsilon_p=c_m\frac{\bm p^2}{2\mu}-c_m^{(2)}\frac{\bm p^{4}}{8\mu^3}+\mO\left(\bm p^{6}/\mu^5\right)\,.
\end{align}
We also introduced four independent coefficients, $b_1$, $b_2$, $b_3$ and $b_4$, given in terms of the Lagrangian parameters in appendix~\ref{AppNREFTscattering}.  One can show that loop corrections do not contribute to the matrix element at this order---see appendix~\ref{SecNREFTLoops}.

{A non-trivial check of our NREFT construction is obtained by comparing the above results to those obtained  in section~\ref{SecMassiveNGBreview} for the benchmark model. Eqs.~\eqref{dispi}, \eqref{eqTripletScattering1} and \eqref{eqTripletScattering2}
should match respectively Eqs.~\eqref{eqNREFTdispersionGappedNGB2}, \eqref{eqNREFTscattering1} and \eqref{eqNREFTscattering2}.
The matching beautifully works, fixing\footnote{ In the matching one must consider that in the triplet model we used the relativistic normalization of states, while in the NREFT (see \eqref{eqQuantizationGappedNGB}) we used the nonrelativistic one
which differs  by a {momentum} dependent factor: $\ket{\bm p,\mu}_\text{triplet}=\sqrt{2 E_p}\ket{\bm p,\mu}_\text{NREFT}$.} }
\begin{align}
\frac{1}{c^{(1)}}=\frac{\lambda \mu^2}{\mu^2-m^2}\,, 
\qquad
&c_m=c_m^{(2)}=1\,,\quad\;
c_s^2=\frac{\mu^2-m^2}{3\mu^2 -m^2}\,,\\
\frac{b_1}{c^{(1)}}=\frac{1}{4}\,,\quad\;
\frac{b_2}{c^{(1)}}=\frac{m^2+\mu^2}{4(m^2-\mu^2)}\,,&\quad\;
\frac{\text{Re}[b_3]}{c^{(1)}}=\frac{7\mu^2+m^2}{4(\mu^2-m^2)}\,,\quad\;
\frac{\text{Re}[b_4]}{c^{(1)}}=\frac{\mu^2}{2(\mu^2-m^2)}\,. \notag
\end{align}
{Notice in particular that the dispersion relation fixes $c_m=1$ at lowest order, which immediately gives Eq.~\eqref{eqNREFTscattering1}
the same momentum dependence as \eqref{eqTripletScattering1}.

This is however not the end of the story. }As already discussed, our benchmark model allows for the process in which two gapped Goldstones annihilate into two gapless ones. {The corresponding amplitude is outside the regime of applicability of the NREFT. Indeed, by energy conservation, the final state consists of modes whose 3-momentum is of the order of the mass of the gapped Goldstones $\mu$, while, as explained, what we have built 
is an EFT valid for processes in which all the external legs are characterized by 3-momenta much smaller than the chemical potential.}
{ Nonetheless, because of the optical theorem, the annihilation rate gives rise} 
to an imaginary part in the {$2\to2$ scattering amplitude of slow gapped Goldstones.}\footnote{The imaginary part induced by elastic scattering itself can be computed within the NREFT and it is of higher order in the velocity.} 
That in turn can be {reproduced} in the NREFT by assigning an imaginary part to the Wilson coefficients, {which then retain some information about the annihilation process. In conclusion, while our EFT is so constructed as to properly realize the symmetry on low momentum amplitudes, the existence of processes that involve large momenta in the final state implies that it cannot be unitary.}

{ We expect  the above statements  to be true in general, for both weakly  and strongly coupled theories. However  an explicit check can only be given  in the former case. Focussing 
on the weakly coupled model of   Sec. \ref{SecMassiveNGBreview},  we will now show that, by
perturbatively matching the UV and IR descriptions,  one does obtain the expected structure of the Wilson coefficients, imaginary parts included. }


Notice first that unitarity of the theory at $\pi=0$ implies that the coefficient $c^{(1)}$ and the sound speed $c_s^2$ of the gapless Goldstone are real---see Eq. \eqref{eqNREFTQuadraticL}. Furthermore,
the accidental $\mathds{Z}_2$ symmetry, which {forbids  gapped Goldstone decay in the linear triplet model}, implies that the coefficients of the dispersion relation \eqref{eqNREFTdispersionGappedNGB2} and, more in general, of all the operators contributing to amplitudes with only one (slow) gapped Goldstone and an arbitrary number of (soft) gapless modes in the initial and final states must be real.\footnote{{This is because, in the linear triplet, the only possible intermediate states contributing to all possible cuts of such amplitudes are those included in the NREFT.}} From inspection of Eqs.~\eqref{eqNREFTLagrangiand3}, \eqref{eqNREFTappB1} and \eqref{eqNREFTappB2}, this implies that $c_m$, $c_m^{(2)}$, $b_1$ and $b_2$ are real as well. Overall, we find that the scattering amplitude must be real at leading order in velocity, while at the subleading order we can use only the imaginary parts of the coefficients $b_3$ and $b_4$ to match the annihilation contribution. {To check  that  is enough}, notice that from
the annihilation cross section~\eqref{eqTripletAnnihilationCrossSection} of the UV theory one finds
\begin{align}\label{eqTripletImaginaryPartScattering}
\text{Im}\left[\mM_\text{elastic}\right]\simeq
\gamma\frac{ (\bm p_a\cdot\bm p_b)^2}{\mu^4}+\delta
\frac{(\bm{p}_a\cdot\bm{p}_c)(\bm{p}_b\cdot\bm{p}_d)
+(\bm{p}_a\cdot\bm{p}_d)(\bm{p}_b\cdot\bm{p}_c)}{\mu^4}\,.
\end{align}
{Non-trivially, this contribution is local and it precisely has the structure to be matched in the NREFT via an imaginary part for $b_3$ and $b_4$}:
\begin{align}
\frac{\text{Im}[b_3]}{[c^{(1)}]^2}=\frac{\gamma}{4}\,,
\qquad
\frac{\text{Im}[b_4]}{[c^{(1)}]^2}=\frac{\delta}{4}\,.
\end{align}


\begin{figure}
\centering
\includegraphics[scale=1.05]{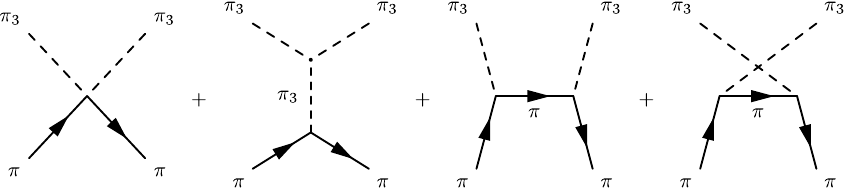}
\caption{Diagrams contributing to $\pi_3\pi\rightarrow\pi_3\pi$ at tree-level.}\label{FigPi3Piscattering}
\end{figure}

\vspace{1em}

{As one last example, to further clarify the procedure of power counting in velocity,
consider the scattering}
$\pi(\bm{p}_a)+\pi_3(\bm{k}_1)\rightarrow 
\pi(\bm{p}_b)+\pi_3(\bm{k}_2)$. The relevant diagrams are presented in figure~\ref{FigPi3Piscattering}. As before, we take all {external 3-momenta  of order  $\mO(\mu v)$. }
One can see that momentum conservation {requires  the intermediate $\pi_3$ of the second diagram to be a potential mode, and the intermediate $\pi$ of the last two to be soft.} One then needs to isolate the relevant interaction terms in the effective Lagrangian, after which it is  straightforward to extract the Feynman rules and compute the matrix elements.

One finds that the leading order result is $\mO(v^3)$ and receives contribution from all diagrams
in figure~\ref{FigPi3Piscattering} but the second, which starts contributing at $\mO(v^4)$. The matrix element reads\footnote{Notice that to leading order in $v$ energy conservation implies $|\bm k_1|=|\bm k_2|$.}
\begin{align}
\begin{split}
\mM&=\frac{1}{2\mu^4c^{(1)}}\frac{1}{c_s|\bm k_1|}
\Big\{(c_s^2-1)c_s^2\bm k^2_1(\bm k_1+\bm k_2)\cdot(\bm p_a+\bm p_b)\\
&\quad+2c_s^2\left[(\bm p_a\cdot\bm k_2)(\bm p_b\cdot\bm k_1)-(\bm p_a\cdot\bm k_1)
(\bm p_b\cdot\bm k_2)
\right]
\Big\}+\mO\left(v^4\right).
\end{split}
\end{align}
This expression vanishes when any of the momenta approaches zero.

A final comment concerns the calculation and power counting of loop diagrams. {As well-known from NRQCD, the formulation of the NREFT at the quantum level is more subtle than in the standard relativistic case, even when using  a mass independent regulator, like dimensional regularization. A consistent treatment, first given in \cite{Gri1,Gri2} and refined in \cite{vNRQCD}, relies crucially on the splitting into soft, potential and ultrasoft modes performed in \eqref{eqSoftUltraSoftPotential}.} The prescription explained there applies straightforwardly to our case. We review some details and provide few examples in appendix \ref{SecNREFTLoops}.

\section{Integrating out the gapped Goldstones: a less effective field theory}\label{secSO(3)=U(1)}

As we already discussed in the Introduction, in {quantum field theory with unbroken Poincar\'e symmetry, the presence of Goldstone modes in the IR has very nontrivial consequences. In particular, 
 Goldstones associated to a} coset $G/H$ signals the existence of a symmetry group $G\times G'$ in the UV. $G$ is spontaneously broken, and $G'$ is any other distinct group which either is trivial or such that all the states charged under it are heavy and absent in the $G/H$ effective theory. This is for instance the case in QCD, where $G=SU(N_f)_L\times SU(N_f)_R$, with $N_f$ the number of light quarks, is broken down to the isospin group $H=SU(N_f)_V$, and the corresponding Goldstones are the light mesons. In this case $G'$ is the baryon number, $U(1)_B$, which is unbroken and whose lightest charged state is the proton.

One might then wonder what happens to our finite density system  when the involved energies, as measured by the unbroken Hamiltonian $\bar{H}=H+\mu Q_3$, are much smaller than the chemical potential $\mu$. One could be tempted to treat the gapped Goldstones just like protons in QCD. {However, while, on the one hand,  they can be integrated out in the EFT at energies $E\ll \mu$, on the other they are needed to  non-linearly realize} the full non-Abelian symmetry. Is there any hint left of the original symmetry once we have integrated them out? In other words, is the information about the non-Abelian nature of the group lost at low energies, similarly {like for   $U(1)_B$  in QCD?}


It is easy to show that in the {zero-$\pi$ sector ($\pi=0$ in the action),} the invariants built out of the coset construction reduce to those of a simple Abelian $U(1)$ group, i.e. $D_\mu\pi=0$ and $D_\mu\chi=\pd_\mu\chi$. It cannot be otherwise, since the internal $SU(2)$ algebra cannot be nontrivially realized on a single field. Physically, when we integrate out the gapped Goldstone we specify  boundary conditions {for it to vanish at infinity. In our case that clearly breaks the non-Abelian symmetry since, as argued in section~\ref{sec:themodel},  symmetry transformations produce a fast oscillating mode that does not decay at infinity.}

That is also evident in the linear triplet model~\eqref{eqTripletLagrangianQuadratic}. At low energies one can, in fact, integrate out explicitly the heavy fields $h(x)$ and $\theta(x)$. At tree level the resulting effective Lagrangian is
\begin{align}\label{eqLeff4}
\begin{split}
\mL_\text{eff}&=\frac{1}{2} \left( 1 + \frac{2\mu^2}{\lambda  \phi_0^2} \right) \dot{\psi}^2 - \frac12 (\bm \nabla\psi)^2 +
\frac{\mu}{\lambda\phi_0^3} \, \dot{\psi}(\pd\psi)^2 + \mO\left(\pd^4/\mu^4\right)\,,
\end{split}
\end{align}
which is a Lagrangian for the Goldstone boson of an ordinary (Abelian) relativistic superfluid, but no other symmetry is manifest.\footnote{In fact, due to the $\mathds{Z}_2$ symmetry, integrating out $\theta$ at tree level accounts to setting it to zero in the Lagrangian~\eqref{eqTripletLagrangianQuadratic}, which turns it into an $O(2)$ doublet theory.} 

\vspace{1em}

To clarify this situation, it is helpful to think in terms of the Hilbert space of the low-energy EFT for the gapless Goldstone only. The latter is obtained by restricting the Hilbert space $\mathcal{H}$ of the full theory to the subspace $\mathcal{H}_\text{EFT}$ specified by the condition:
\begin{align}\label{eqEFT_Hilbert}
\ket{\psi}\in\mathcal{H}_\text{EFT}\quad\iff\quad\langle \psi|\bar{H}|\psi\rangle=\langle \psi|H+\mu Q_3|\psi\rangle\ll \mu\,.
\end{align}
Despite the theory being $SU(2)$ invariant, the presence of $Q_3$ in the modified Hamiltonian that we use to specify the configurations that are part of the EFT explicitly breaks the symmetry. As a concrete illustration, consider a free quantum mechanical particle living on a sphere, with Lagrangian $L=\frac{I}{2}\left(\dot{\theta}^2+\sin^2\theta\,\dot{\phi}^2\right)$, where $I$ is the moment of inertia. The states of the theory are organized in $SO(3)$ multiplets, $\ket{\ell,m}$, with energy $E_{\ell}=\ell(\ell+1)/2I$. The quantum number $\ell$ specifies the representation and $m$ is the value of the angular momentum along the $z$-axis: $-\ell\leq m\leq\ell$. If we take $m$ to be fixed, negative and large, the state with minimum energy is $\ket{\ell,m=-\ell}$ and the chemical potential is $\mu=\pd E_{\ell=-m}/\pd m\approx m/I$ \cite{MoninCFT}; any other state in the same $SO(3)$ multiplet has a gap of at least $|\mu|\approx|m|/I$ as measured by $\bar{H}$. Thus, for every fixed value of the third component of the angular momentum, the low-energy EFT is made of the single state $\ket{\ell,m=-\ell}$, which is not invariant under the full rotation group. At the Lagrangian level, {the restriction to such states corresponds to ``integrating out" the polar angle, $\theta$, considering an effective theory for the azimuthal angle, $\phi$, spinning around the $z$-axis.} Indeed, a single excitation of $\theta$ describes a state with total angular momentum increased by a unity, $\ell+1$, but with the same projection along the $z$-axis, $m=-\ell$. This corresponds to a state with gap $|\mu|$ at large angular momentum~\cite{MoninCFT}. This is analogous to the gapped Goldstone, providing a simple illustration of its key role in the nonlinear realization of the full symmetry group.\footnote{ In field theory (at infinite volume) the action of the spontaneously broken charges on the Hilbert space of the theory is not well-defined and we cannot classify state according to representation of the broken group; this however does not invalidate our main point, that the restriction \eqref{eqEFT_Hilbert} \emph{explicitly} breaks  the symmetry. }


\vspace{1em}

The condition in Eq.~\eqref{eqEFT_Hilbert} implies that the theory without the gapped Goldstone can only be used to compute correlators whose long-distance behaviour is determined by intermediate states with small energy under $\bar{H}$. However, since time evolution is still controlled by the Hamiltonian $H$, not all correlation functions having a non-trivial long-distance limit satisfy this property. In other words, the operators corresponding to such correlation function cannot be matched in the low-energy EFT for the gapless Goldstone only, and they would simply be lost. In contrast, if one employs the NREFT we described so far, the previous correlators can be consistently reproduced within its regime of applicability. As an illustration, consider the {time} component of the Noether currents for the $Q_+$ and $Q_-$ generators of $SU(2)$. It is clear that, in an EFT that only contains the gapless Goldstones, such operators cannot be matched. Indeed, in  such a theory, only the Abelian subgroup of $SU(2)$ is realized nontrivially, and the Noether currents associated to $Q_\pm$ cannot be computed. 
On the other hand, working in the NREFT, in which the full non-Abelian symmetry group is realized, it is straightforward to compute them from Noether theorem and, at leading order in fields and derivatives, we find
\begin{align}\label{eq_JJ_H}
J^0_-(t,\bm{x})\simeq-ic^{(1)}\mu^3\pi(t,\bm{x})\,,\qquad
J^0_+(t,\bm{x})\simeq ic^{(1)}\mu^3\pi^*(t,\bm{x})\,.
\end{align}
As it could have been expected from the conservation of the global charges, these are written purely in terms of the \emph{slow} field $\pi$ of the Right parametrization~\eqref{eq:coset2}. We can now compute their correlators at large time separation and spatial distance. For instance, the spatial Fourier transform of the two-point function of these currents can be computed from the gapped Goldstone propagator and reads
\begin{align}\label{eq_JJ_correlator}
\int d^3x\, e^{-i\bm{p}\cdot\bm{x}}\braket{\mu|T\left\{J^0_-(t,\bm{x})J^0_+(0,\bm{0})\right\}|\mu}
=2c^{(1)}\mu^3\theta(t)e^{-i\epsilon_p t}\,,
\end{align}
where $T$ is the time-ordered product and $\epsilon_p$ is the (possibly complex) kinetic energy of the gapped Goldstone, given by Eq.~\eqref{eqDispersionGappedNGB} at leading order in  {3-momentum}. 
For long wavelengths, $|\bm{p}|\ll\mu$, the correlator \eqref{eq_JJ_correlator} oscillates slowly in time---i.e. it has nontrivial long time tails. Nonetheless, it cannot be computed from the low-energy EFT without the gapped Goldstone, as already anticipated.\footnote{That this result cannot be obtained by somehow matching the currents in the low-energy theory is also manifest from the fact that the correlator oscillates with frequency $\epsilon_p\sim\bm{p}^2/\mu$, while no state with such dispersion relation is present in the EFT for the gapless Goldstone only.} This is clear when the gapped Goldstone is stable and $\epsilon_p$ is real, in which case the result in Eq.~\eqref{eq_JJ_correlator} is interpreted as the free evolution in time of a single $\pi$ mode. Such a simple interpretation does not exist in more general cases, but this does not affect the main picture presented above.\footnote{Equivalently, one could look at the operator $\bar{J}_\pm^0(t,\bm{x})\equiv e^{i\bar Ht} J_\pm^0(0,\bm{x})e^{-i\bar Ht}$, which instead evolves with $\bar H$. It is simple to show that the two-point correlator for this (non-conserved) current oscillates with frequency $\mu$. Consequently, it can never be obtained from the EFT for the gapless Goldstones only, which has support only on frequencies $\ll \mu$, as measured by $\bar{H}$.}




\vspace*{1em}

In summary, in the low-energy EFT specified by Eq.~\eqref{eqEFT_Hilbert} no signature of the non-Abelian nature of the symmetry is present. To obtain a fully $SU(2)$ covariant description one should work within the NREFT presented in this work, which reduces to the Abelian superfluid in the zero gapped Goldstone sector. {In particular, our construction shows that the non-Abelian structure of the group constrains the dynamics at small \emph{spatial} momenta, similarly to the relativistic case, but around non-zero frequencies which are multiples of the chemical potential.} The NREFT further provides access to certain non-trivial correlation functions at large spacetime separations, which cannot be matched without the gapped Goldstone due to the difference between the fundamental Hamiltonian $H$ and $\bar{H}$. We illustrated that point by discussing the two-point function of the $SU(2)$ Noether current; we leave a systematic analysis of operator matching in the NREFT for future work. 
These considerations, we believe, clarify previous works \cite{MoninCFT,Bern1,HellermanO41}, which, at large chemical potential, restricted their attention to the Abelian component of the spontaneously broken internal symmetry. 
We conclude this section marking the differences between {the present case and the relativistic case, i.e. a broken internal symmetry with unbroken Poincar\`e invariance.

In the relativistic case symmetry constrains all the Goldstone bosons to have 4-momentum on the lightcone. Then, given a coset $G/H$, the gapless Goldstone bosons carry all the information about the symmetry breaking and, as made evident by the CCWZ construction, all degrees of freedom falling into gapped $H$-multiplets can be integrated out preserving the full $G$ symmetry.
As concerns instead the role of an additional unbroken $G'$ factor in the fundamental symmetry, if all the  states charged over $G'$ are gapped, then the corresponding Noether currents do not have low frequency components. In view of that in no way the low energy modes can match them, and the information about $G'$ is lost in the EFT. {A similar situation arises for gapped Goldstones, which cannot be integrated out while still preserving the full $G$ symmetry. However, in this case
 the currents that interpolate for the gapped Goldstones do have low frequency components---see Eq.~\eqref{eq_JJ_correlator}---}and there must therefore exist a way to recover that information via an EFT construction, ours indeed.}

%

\section{Conclusions and future directions}

{
The breaking of internal symmetries has qualitatively different implications on  low-energy physics, depending on whether or not it is accompanied by the breaking of spacetime symmetries. One crucial difference arises for the spectrum of excitations. With unbroken Poincar\`e invariance, Goldstone theorem dictates the presence of one stable particle with light-like dispersion relation, $E(\bm{k})=|\bm{k}|$, for each spontaneously broken symmetry generator. With the spontaneous breaking of the Poincar\`e group, Goldstone theorem
leaves instead space for a greater  variety of options,  as concerns the counting of modes,  their dispersion relations and their stability. A particularly interesting case is offered by non-Abelian  superfluids, which are characterized by chemical potentials $\mu_I$ for the Cartan charges $Q_I$. Here  Goldstone theorem implies the presence of a set of modes, labeled by  $a=1,\dots,N$, whose energy satisfies $E_a(\bm{k} = 0) =c_{aI}\mu_I$, with $c_{aI}$ real coefficients that are fully dictated by group theory~\cite{Nicolis_Theorem}. Generically one then has both gapless modes, $E_a(0) =0$ and gapped ones $E_a(0)\not = 0$. Moreover one has variety in the functional dependence of $E_a(\bm{k})$  on  
$\bm k$, including the possibility for imaginary parts, associated, when allowed,  with the decay of the modes at $\bm{k}\not = 0$.

Symmetry controls not only the spectrum, but also the  interaction of the Goldstone bosons. In the Poincaré invariant case,  this results in a low-energy EFT whose main features are 
 {\it universal} and rather independent of the details of the microphysics.
In finite density systems constraints on the structure of the interactions are expected, and, to some extent, have been studied. However, with gapped Goldstones, the EFT construction also raises issues of technical and conceptual nature. One  concerns universality, and stems from the  generic possibility
of other, non-Goldstone degrees of freedom in the range of energies and momenta  $\mathcal{O}(\mu)$. Those are, for instance, expected in systems like CFTs,  where $\mu$ is the main dimensionful parameter. In that situation creation and destruction of gapped Goldstones, even slow moving ones, entails momenta $\sim \mu$  evading a universal EFT description.
Another issue concerns the possibility of reconstructing the pattern of symmetry breaking by pure consideration of the dynamics at the lowest possible energies. That is possible in the relativistic case, but seems impossible at finite density, as the gapped Goldstones are integrated out at $E\ll \mu$.\footnote{An interesting question regards whether gapped Goldstones can be excited by some light external probe charged under the internal symmetry. We leave this investigation for future work.}

In this paper we have clarified the above questions.  We have shown that the EFT that universally implements the {information on  the symmetry breaking pattern has degrees of freedom given by the Goldstone modes, all of them, at low 3-momentum, $\bm k$}. In particular the gapped Goldstones are limited to small velocity, which  also manifestly controls the strength of their interactions, in agreement with~\cite{Brauner}.
Such EFT cannot produce  amplitudes that violate gapped Goldstone number (GGN), as these necessarily involve external legs  with large 3-momentum $\sim \mu$. Consequently GGN is an  ``emergent" symmetry of the EFT where time evolution proceeds without transitions between Hilbert spaces with different GGN. This  bars the calculability of physical processes where the GGN is not conserved. The latter are nonetheless consistently described in an {\it inclusive} form through the optical theorem, by allowing for imaginary parts in the local coefficients of operators in the EFT. The price to pay is that the unitarity of the original theory is not manifest in the EFT. The fact 
that GGN non-conservation involves short wavelength modes however allows to describe it via local operators in the EFT. The resulting picture is fully analogous to that of non-relativistic EFTs (NREFTs), like for instance non-relativistic QCD \cite{NRQCD,vNRQCD} or the EFT for nucleon-nucleon scattering \cite{Weinberg:1990rz,KaplanEFT}, which have indeed almost completely guided our construction. We have illustrated our ideas by focussing on an $SU(2)$ superfluid, where we also checked that the results of the EFT construction match those of an explicit renormalizable model. We expect our results to be easily generalizable to arbitrary symmetry breaking patterns, as well as to allow the inclusion of other possible relevant matter fields in the action via standard techniques~\cite{Weinberg2}.

With the above  picture in place it is evident that the complete information about symmetry breaking in the microscopic theory is encoded in the  full set of NREFTs Hilbert spaces with 
all possible GGN. The subspace with zero GGN, which purely involves the soft gapless modes, is only part of the picture and does not encode the complete information about symmetry breaking. In particular it does not contain information about the spectrum of gapped modes. This subspace also happens to correspond to the EFT describing the lowest lying modes 
of the unbroken time translation generator $\bar H=H+\mu_I Q_I$ of the superfluid. This Hamiltonian is  only  invariant under a subgroup of the original 
internal symmetry, which makes it clear why such lowest energy EFT cannot describe the full pattern of symmetry breaking. A more detailed discussion of this is given in section \ref{secSO(3)=U(1)}.}

\vspace{2em}

Before closing let us discuss a few possible applications of the gapped Goldstone NREFT.
As remarked in the introduction, gapped Goldstones appear in different physical systems~\cite{WatanabeMNGB}. An interesting example is given by QCD at finite density, as it is for instance found in the interior of neutron stars~\cite{Kaplan_KaonCondensate,Son_Kaon1,Son_Kaon2,Brown_NeutronStar}. Depending on the parameters, in particular baryon density, it is
conceivable that  the system  relaxes to a superfluid phase for the  non-abelian isospin symmetry. One concrete possibility is represented by Kaon condensation~\cite{Kaplan_KaonCondensate}.
The resulting scenario, given the approximate nature of the isospin symmetry,  broken by the small quark masses, would be approximated by the physical situation described in this paper: there would be pseudo-Goldstone bosons, whose gap and interactions are controlled by symmetry breaking, spontaneous and explicit, very much like in the QCD chiral Lagrangian around the vacuum.
 In particular in the regime where the chemical potential is of the order of the strong interaction scale, our NREFT would capture, amid a hardly calculable strong  dynamics, the universal features of the gapped pseudo-Goldstones dynamics.

The underlying Lorentz invariance of the theory, if conceptually useful in understanding the origin of the modified Hamiltonian $\bar{H}$, is not necessary for the existence of both gapless and gapped Goldstone bosons \cite{WatanabeMNGB}. Indeed, our construction may be straightforwardly applied to systems where either only the Galilean limit of Lorentz transformations is considered, or boost invariance is not present from the beginning.\footnote{Physically, this means that boost invariance is broken by some more microscopic dynamics, typically due to the presence of a lattice or some other fluid, whose associated 
hydrodynamics modes can be neglected in first approximation.}
 Possibly relevant examples of this kind include ferromagnets,  anti-ferromagnets \cite{Leutwyler},  electron gases \cite{Kohn} and vortex lattices \cite{Moroz1} where spin or angular momentum play the role of the non-Abelian charges, while the role of the chemical potential is played by  either a uniform magnetic field \cite{WatanabeMNGB} or by an externally induced angular velocity.
  In these examples the role of the gapped Goldstones is played 
 respectively by the magnons for spin systems and by the Kohn mode for electron gases and vortex lattices. It would be interesting to investigate the possibility to apply our NREFT methodology to such systems, searching in particular for situations where the Goldstone gap is comparable to or larger than the energy of other potentially strongly coupled modes. Our methodology would allow to zoom on the universal properties of otherwise hardly tractable strongly coupled systems.

\vspace{1em}

Recently, effective field theory {techniques} have been applied in the study of
large charge operators in conformal field theories~\cite{Hellerman,MoninCFT,Bern1,HellermanO41}. {By the state/operator correspondence, these are associated {with} condensed matter phases~\cite{Nicolis_Zoology,Son:2005rv,RothsteinFL}, with  the generalized superfluid  described in this paper representing the simplest possibility.}
The NREFT discussed here, when specialized to the cylinder, is then expected to apply in the large charge sector of CFTs invariant under non-Abelian symmetry groups. 

{Interestingly, we can learn something about the spectrum of the strongly interacting conformal $O(3)$ model, using some inputs from the study of the linear triplet model in section}~\ref{SecMassiveNGBreview}. Indeed, since the triplet describes the $O(3)$ Wilson-Fisher fixed point in $4-\varepsilon$ dimension, we expect the large charge sector of the related $3d$ CFT to undergo the same symmetry breaking pattern:\footnote{Where symmetry breaking is intended in the sense explained in \cite{MoninCFT}.} $O(3)\rightarrow\mathds{Z}_2$. The $\mathds{Z}_2$ crucially implies that single gapped Goldstone states, being charged under the latter, are exact eigenstates of the Hamiltonian. {They are thus stable  in the infinite volume limit}.
 By the state-operator correspondence, {they} are associated with $\mathds{Z}_2$ odd operators of angular momentum $J$ transforming in the $(2Q+1)$-representation of the internal $SO(3)$ in the corresponding three-dimensional CFT; the NREFT then allows to compute their scaling dimension as
\begin{align}
\Delta^{(J)}_{mNGB}(Q)=\Delta_0(Q)+\mu(Q)+c_m\frac{J(J+1)}{2\mu(Q)}+
\mO\left(\frac{J^4}{\mu^3(Q)}\right),
\end{align}
where $\Delta_0(Q)$ is the scaling dimension of the lightest scalar operator in the $SO(3)$ $(2Q+1)$ representation, given in a large $Q$ expansion by
\begin{align}
\Delta_0(Q)=\alpha Q^{3/2}+\beta Q^{1/2} - 0.0937256 +\gamma Q^{-1/2}
+\mO\left(Q^{-1}\right),
\end{align}
while $\mu=\pd \Delta_0(Q)/\pd Q\sim Q^{1/2}$ is the chemical potential in units of the cylinder radius and $\alpha$, $\beta$, $\gamma$ and $c_m$ are Wilson coefficients. As in the Abelian case, also massless phonon states {correspond} to CFT operators 
\cite{Hellerman,MoninCFT}.

{Notice that  in a general  $SU(2)$ invariant CFT,  there is no conserved $\mathds{Z}_2$ and things are made more involved by the mixing of the gapped Goldstone with states made out of lighter particles, {outside the validity of the NREFT}. Such mixing however corresponds to the decay of the gapped Goldstone state in the infinite volume limit.
Therefore  the NREFT approach should allow the description of the resulting {\it inclusive} features,  presumably encoded in the  spectral distribution. Relatedly,  the 
NREFT should allow to  match all the components of the non-Abelian Noether current in terms of Goldstone fields, in a certain kinematic regime. We plan to investigate the detailed predictions of the NREFT for {CFTs with non-Abelian symmetry}  in a future work.}

\section*{Acknowledgements}

We thank T.~Brauner, I.~Low, R.~Penco, J.~Penedones, F.~Piazza, I.~Rothstein and M.~Stephanov for useful discussions, and especially A.~Nicolis for collaboration at the early stages of this work.
This work has been supported by the Swiss National Science Foundation under contract 200020-169696 and through the National Center
of Competence in Research SwissMAP. E.G. acknowledges support by the Deutsche Forschungsgemeinschaft (DFG, German Research Foundation) under Germany’s Excellence Strategy -- EXC 2121 “Quantum Universe” -- 390833306.

\appendix

\section{Triplet model details}\label{AppTriplet}

\subsection{Coefficients of the gapped Goldstone annihilation}\label{AppTripletAnnihilation}

The coefficients in \eqref{eqTripletAnnihilatioMatrixElement} are given by
\begin{align*}
\alpha&=\frac{  2 \left(5 \mu ^2-3 m^2-\sqrt{4 \mu ^4+\left(\mu^2-m^2\right)^2}\right)}{ 2 \mu ^2 + \sqrt{4 \mu ^4+\left(\mu^2-m^2\right)^2}}\,, \\
\beta &=\frac{-8 \, \mu ^2 \left(\mu ^2-m^2\right)^{2}}{29 \mu ^6-m^2\mu ^4+3m^4\mu^2 -m^6+\left(13 \mu ^4+2 m^2\mu ^2 
+m^4\right) \sqrt{4 \mu ^4+\left(\mu^2-m^2\right)^2}}\,.
\end{align*}
Those in \eqref{eqTripletAnnihilationCrossSection} are
\begin{align*}			
\gamma&=\frac{ \lambda^2 \mu ^3
\left[\frac{\sqrt{5 \mu ^4+m^4-2 m^2 \mu ^2 }+m^2}{5 \mu ^4+m^4-2 m^2  \mu ^2 }\right]^{1/2} }{15 \pi  \left(\mu ^2-m^2\right)^6}\times\\ \nonumber
			&\quad \times\Big[2085 \mu ^{10}-49 m^{10}+441 m^8 \mu ^2 -1762 m^6 \mu ^4  +3842 m^4 \mu ^6 -4429 m^2  \mu ^8 +\\ 
			&\quad -\left(935 \mu ^8+55 m^8-432 m^6 \mu ^2 +1314 m^4 \mu ^4 -1808 
			 \mu ^6 m^2\right) \sqrt{5 \mu ^4+m^4-2m^2  \mu^2}\Big]\,, \notag \\
			\delta&=\frac{-2 \lambda ^2 \mu ^2 \left(\mu ^2-m^2\right)^2 \left(2 \mu +\frac{4 \mu ^3}{\sqrt{5 \mu ^4+m^4-2 m^2  \mu ^2 }}\right) \left(\sqrt{5 \mu ^4+m^4-2 m^2  \mu ^2 }+m^2\right)^{5/2}}{15 \pi  \left[29 \mu ^6+m^6+3 \mu ^2 m^4-m^2 \mu ^4 +\left(13 \mu ^4+m^4+2 m^2  \mu ^2 \right) \sqrt{5 \mu ^4+m^4 -2 
			 \mu ^2 m^2}\right]^2}\,.
		\end{align*}

\subsection{Gapped Goldstone decay}\label{AppTripletDecay}

In the linear triplet model discussed in the main text, the accidental discrete $\mathds{Z}_2$ symmetry forbids the decay of the gapped Goldstone. However, in more general theories the gapped Goldstone can decay into arbitrary lighter states. Here we provide a simple example of such a modification of the Lagrangian~\eqref{eqTripletLagrangian}. The resulting decay rate vanishes with the 3-momentum of the gapped Goldstone, in agreement with the general discussion of section~\ref{sec:themodel}.

To induce a decay channel for $\theta$, we need to break explicitly the $\mathds{Z}_2$ symmetry of the Lagrangian \eqref{eqTripletLagrangian}. In order to do that, we couple the $O(3)$ triplet $\bm{\Phi}$ to a complex $U(2)$ doublet $ \Psi$. We hence add the following term to the linear triplet model Lagrangian:
\begin{align}
\delta\mL=\left|\pd\Psi\right|^2-m_{\Psi}^2\left|\Psi\right|^2-\frac{\lambda_\Psi}{4}
\left|\Psi\right|^4-g\left(\Psi^\dagger\frac{\bm{\sigma}}{2}\Psi\right)\cdot\bm{\Phi}-\frac{\gamma}{4}\left|\Psi\right|^2\bm{\Phi}^2\,.
\end{align}
Here $\bm{\sigma}=\left(\sigma_1,\sigma_2,\sigma_3\right)$ are the Pauli matrices. Adding this term to \eqref{eqTripletLagrangian}, the resulting Lagrangian is the most general renormalizable theory of a doublet and a triplet preserving a global $SU(2)\times U(1)$ symmetry.
Crucially, the coupling $g$ breaks the discrete $\mathds{Z}_2$ symmetry which prevented $\theta$ from decaying.  
All parameters are positive. When not specified otherwise, all parameters with the same coupling and mass dimensions are assumed to be of the same order \cite{RiccardoSILH}.

We expand around the VEV \eqref{eqTripletVEV} for the triplet with $\Psi=0$, which is a minimum for 
\begin{align}\label{eqTripletDoubletCondition1}
\gamma\geq 2\frac{g}{\phi_0}+\frac{\mu^2-4m_{\Psi}^2}{\phi_0^2}\,.
\end{align}
This leaves the $U(1)$ acting as $\Psi\mapsto e^{i\alpha}\Psi$ unbroken.
In this case the fluctuations for $\bm{\Phi}$ are parametrized as before (see Eq.~\eqref{eqTripletFluctuations}) while $\Psi$ can be written as
\begin{align}
\Psi=e^{-i\left(\mu t+\psi(x)/\phi_0\right)\frac{\sigma_3}{2}}
\left(\begin{array}{c}
\Psi_1(x) \\
\Psi_2(x)
\end{array}\right),\qquad
\Psi_1,\;\Psi_2\;\in\mathds{C}.
\end{align}
Notice that we explicitly factored out a time dependent rotation, which makes explicit that unbroken time translations correspond to $H+\mu Q_3$.
To find the spectrum, consider the quadratic contribution from $\delta\mL$:
\begin{align}
\begin{split}
\delta\mL^{(2)}&=|\pd \Psi_1|^2+|\pd \Psi_2|^2+\frac{1}{2}i\mu\left(\Psi_1^*\dot{\Psi}_1-\Psi_2^*\dot{\Psi}_2-\text{c.c.}\right)
-\frac{g}{2}\phi_0\left(\Psi_1^*\Psi_2+\text{c.c.}\right)\\
&\quad-\left[m_{\Psi}^2+ \frac{\gamma}{4\lambda}m^2
+\left(\gamma/\lambda-1\right)\mu^2/4\right]\left(|\Psi_1|^2+|\Psi_2|^2\right)\,.
\end{split}
\end{align}
The fields $\Psi_1$ and $\Psi_2$ interpolate four quasi-particles: $\{\ket{\Psi_+(\bm{k})},\ket{\Psi_-(\bm{k})},\ket{\bar{\Psi}_+(\bm{k})}$, $\ket{\bar{\Psi}_-(\bm{k})}\}$. Under the unbroken $U(1)$, $\ket{\Psi_\pm(\bm{k})}$ have positive charge while $\ket{\bar{\Psi}_\pm(\bm{k})}$ have negative charge. As a consequence of the symmetry $\Psi_1\leftrightarrow \Psi_2^*$ of the quadratic Lagrangian, oppositely charged modes have dispersion relations equal in pair, given by:
\begin{align}
\omega^2_{\pm}(k)=\bar{\omega}^2_{\pm}(k)=
\frac{\mu ^2}{4}+m_{\Psi}^2+\frac{\gamma  }{4}\phi_0^2+
k^2
\pm\sqrt{\frac{\gamma  \mu ^2}{4 }\phi_0^2+\frac{g^2 }{4 }\phi_0^2
+\mu ^2 m_{\Psi}^2+k^2 \mu ^2}\,.
\end{align}
Here $\omega_{+}(k)=\bar{\omega}_+(k)$ is the dispersion relation of 
$\ket{\Psi_+(\bm{k})}$ and $\ket{\bar{\Psi}_+(\bm{k})}$, while 
$\omega_{-}(k)=\bar{\omega}_-(k)$ is the dispersion relation of $\ket{\Psi_-(\bm{k})}$ and $\ket{\bar{\Psi}_-(\bm{k})}$.
Notice further that, because of the aforementioned symmetry of the quadratic Lagrangian, the wavefunctions of the fields on the states $\{\ket{\Psi_-(\bm{k})},\ket{\bar{\Psi}_-(\bm{k})}\}$ satisfy
\begin{align}\label{eqTripletDoubletMatrixElements}
\braket{0|\Psi_{1/2}(0)|\Psi_-(\bm{k})}=
e^{i\alpha}\braket{0|\Psi^*_{2/1}(0)|\bar{\Psi}_-(\bm{k})}\,,\qquad
\alpha\in\mathds{R}
\end{align}
where $e^{i\alpha}$ is an unphysical phase factor which depends upon the precise definition of the states $\ket{\Psi_-(\bm{k})}$ and $\ket{\bar{\Psi}_-(\bm{k})}$. We will use this relation in the following.

The gapped Goldstone couples linearly to the complex $U(2)$ doublet through the $\mathds{Z}_2$ breaking coupling $g$:
\begin{align}\label{eqTripletDoubletVertex}
-g\left(\Psi^\dagger\frac{\bm{\sigma}}{2}\Psi\right)\cdot\bm{\Phi}
\supset\frac{g}{2}\theta\left(|\Psi_2|^2-|\Psi_1|^2\right)\,,
\end{align}
To induce a decay for $\theta$, we need the gap of the modes $\{\ket{\Psi_-(\bm{k})},\ket{\bar{\Psi}_-(\bm{k})}\}$ to be less than half of the gapped Goldstone mass: $\omega_-(0)=\bar{\omega}_-(0)\leq\mu/2$. This happens for\footnote{The conditions \eqref{eqTripletDoubletCondition1} and \eqref{eqTripletDoubletCondition2} are compatible, as it can be seen in the limit where $\mu$ is much bigger than all other mass parameters where they reduce to $\lambda\leq \gamma\leq 4\lambda$.}
\begin{align}\label{eqTripletDoubletCondition2}
m_{\Psi}^2+\frac{\gamma  }{4}\phi_0^2
-\sqrt{\frac{\gamma  \mu ^2}{4 }\phi_0^2+\frac{g^2 }{4 }\phi_0^2
+\mu ^2 m_{\Psi}^2}\;\leq 0\,.
\end{align}
Under this condition, the following decay channel exists for $\theta$
\begin{align}
\theta(\bm p)\rightarrow\Psi_-(\bm{k}_1)+\bar{\Psi}_-(\bm{k}_2)\,.
\end{align}
It is easy to compute the associated matrix element induced by the vertex \eqref{eqTripletDoubletVertex}; we do not report the details of the calculation. Notice however that the relation \eqref{eqTripletDoubletMatrixElements} implies that the decay amplitude vanishes when the final states have the same momenta. Consequently, a gapped Goldstone at rest cannot decay, as expected. Noticing that $|\bm{k}_1|$ is generically of order $\mO(\mu)$, to linear order in the velocity the matrix element reads
\begin{align}
i\mM=i C \frac{\bm{p}\cdot\bm{k}_1}{|\bm{k_1}|}+\mO\left(\bm{p}^2/\mu,(\bm{p}\cdot\bm{k}_1)^2/\mu^3\right),
\end{align}
where $C$ is
\begin{align}\nonumber
C&=\frac{g^2\mu\phi_0/2}{
2 \mu ^2 \left(\sqrt{g^2 \phi_0^2+\mu ^4}+\mu ^2\right)+g^2 \phi_0^2}
\\
&\quad\times \sqrt{\frac{   \left(2 \sqrt{g^2 \phi_0^2+\mu ^4}
+2 \mu ^2-\gamma  \phi_0^2-4 m_{\Psi}^2\right)}
{ 3 \mu^2+2 \sqrt{g^2 \phi_0^2+\mu ^4}-2 \sqrt{2 \mu ^2 \left(\sqrt{g^2 \phi_0^2+\mu ^4}+\mu ^2\right)+g^2 \phi_0^2}}}\,.
\end{align}
In the limit where $\mu$ is much bigger than all other mass parameters this expression simplifies to
\begin{align}
C=\frac{g^2 \sqrt{4 \lambda-\gamma }}{8 \lambda \mu ^2}+\mO\left(\mu^{-4}\right)\,.
\end{align}
The total decay rate finally takes the following simple form
\begin{align}
\Gamma=c\frac{\bm{p}^2}{\mu}=\left[\frac{g^4  (4 \lambda-\gamma )^{3/2}}{1536 \pi  \lambda^{5/2} \mu ^4}+\mO\left(\mu^{-6}\right)\right]\frac{\bm{p}^2}{\mu}\,,
\end{align}
where $c$ is a dimensionless constant which we wrote in the $\mu\rightarrow\infty$ limit for illustration in the right hand side.

\section{The spacetime coset construction} \label{sec:cosetgeneral}

In this section we review the standard coset construction in presence of broken spacetime symmetries. Our goal is to show how to recover the Lagrangian in Eq. \eqref{eq_GenLag} from this approach. Furthermore, this construction provides a useful bookkeeping tool to build higher derivative terms in our action, which we do in appendix \ref{AppNREFTAction}.

Consider a relativistic system with an internal $SU(2)$ symmetry, whose charge $Q_3$ is at finite density. The ground state $|\mu\rangle$ of such a system minimizes the modified Hamiltonian $\bar H= H+\mu Q_3$~\cite{Nicolis_SSP}, and it can be chosen to satisfy\footnote{In general, the ground state will satisfy $\bar H|\mu\rangle=\lambda |\mu\rangle$, with minimum $\lambda$. In the absence of gravity, one can always add a cosmological constant term to the Hamiltonian to set $\lambda=0$, with no physical consequences~\cite{Nicolis_SSP}.}
\begin{align}\label{eqHeffMu}
\bar H\ket{\mu}=\left(H+\mu Q_3\right)\ket{\mu}=0\,.
\end{align}
If $Q_3$ is spontaneously broken so is $H$, the generator of time translations. The generators of boosts, $J_{0i}$, and the other internal generators, $Q_1$ and $Q_2$, are broken too. The symmetry breaking pattern is then
\begin{align}
\begin{split}
\text{unbroken}&=
\begin{cases}
\bar{H}=H+\,\mu Q_3  
&\text{time translations}\,,\\
\bar{P}_i=P_i
&\text{space translations}\,,\\
J_{ij}
&\text{rotations}\,,
\end{cases}
\\ 
\text{broken}&=
\begin{cases}
J_{0i}\hphantom{=P_0+\,\mu Q_3\,}
& \text{boosts}\,, \\
Q_3,Q_1,Q_2
& \text{internal symmetries}\,. \\
\end{cases}
\end{split}
\end{align}
Therefore we have a theory with a symmetry group, $G$, given by the product of Poincar\'e and the internal $SU(2)$, which is spontaneously broken down to the semidirect product of the modified translations, generated by $\bar{P}_\mu=\{\bar H,\bar{\bm P}\}$, and rotations. We denote the unbroken group with $G^\prime$. Following the standard CCWZ procedure, the coset $G/G'$ can be parametrized as
\begin{align}\label{eqCosetFull}
\Omega=e^{i \bar{P}_\mu x^\mu}e^{i\eta^i J_{0i}}e^{i\pi_3 Q_3}e^{i\alpha \frac{Q_+}{2}+i\alpha^* \frac{Q_-}{2}}\,.
\end{align}
The way to construct an action which is invariant under the full symmetry group is to consider the Maurer-Cartan form, $\Omega^{-1}d\Omega$, and expand it in the basis of broken and unbroken generators. Its general expression reads
\begin{align}\label{eqMC1formFull}
\Omega^{-1}\pd_\mu\Omega=i e^{\;\;a}_\mu\left(\bar{P}_a
+\nabla_a\eta^i J_{0i}
+\nabla_a\pi_3 Q_3+\nabla_a\alpha\frac{Q_+}{2}+\nabla_a\alpha^*\frac{Q_-}{2}+\frac{1}{2} \omega_{a}^{ij}J_{ij}\right).
\end{align}
Here $e^{\;\;a}_\mu$ transforms as a spacetime vielbein~\cite{IvanovGravity1,MoninWheel}, and we introduced Latin indices $a,b=0,1,2,3$ and $i,j=1,2,3$ to distinguish within the vielbein indices, as in the familiar geometrical case. The coefficients of the broken generators, $\nabla_a \eta^i$, 
$\nabla_a\pi_3$ and $\nabla_a\alpha$, are the covariant derivatives of the Goldstones. They have the property that, under the action of any element of the full group, they transform as a linear representation of the \emph{unbroken} subgroup. Finally, $\omega_a^{ij}$ transforms as a spin connection~\cite{ogievetsky1974nonlinear}, which can be used to build higher covariant derivatives of the Goldstone fields:
\begin{align} \label{eqCovariantCCWZderivative}
\nabla_a^H=e^{\;\;\mu}_a \partial_\mu+\frac{i}{2}\omega^{ij}_aJ_{ij}\,.
\end{align}
The previous derivative can also act on additional matter fields that transform in some linear representation of the unbroken group $G^\prime$. 
The most general Lagrangian for the Goldstones, which is invariant under nonlinearly realized symmetry $G$ is then given by
\begin{align} \label{Leffstandard}
\mathcal{L}_\text{eff}=F(\nabla_a \Psi,\nabla_a^H\nabla_b\Psi,\dots)\,,
\end{align}
where we have collectively represented the Goldstone fields as $\Psi$. Here $F$ is any function that depends on combinations of its arguments that are manifestly invariant under the unbroken group.\footnote{In this case, this just means that space indices $i,j,\ldots$ should be contracted in a rotationally invariant way.}

For the case at hand, let us define $(e^{-i\eta^i J_{0i}})_{\;\;\mu}^{a}=(\Lambda^{-1})_{\;\;\mu}^a=\Lambda_\mu^{\;\;a}$ and $\chi=\mu t+\pi_3$ \cite{MoninCFT}. The quantities defined in \eqref{eqMC1formFull} then read
\begin{gather}\label{eq:MCders}
\begin{split}
e_\mu^{\;\;a}=\Lambda_\mu^{\;\;a}\,,&\qquad
\nabla_a\eta^i=-\Lambda^\mu_{\;\;a}(\Lambda^{-1}\pd_\mu\Lambda)^{0i}\,,\qquad
\omega_{a}^{ij}=-\Lambda^\mu_{\;\;a}(\Lambda^{-1}\pd_\mu\Lambda)^{ij}\,,\\
&\nabla_a\pi_3=\Lambda^\mu_{\;\;a}D_\mu\chi-\mu\delta_a^0\,,\qquad
\nabla_a\alpha=\Lambda^\mu_{\;\;a}D_\mu\alpha\,,
\end{split}
\end{gather}
where $D_\mu\alpha$ and $D_\mu\chi$ are the covariant derivatives for a Lorentz invariant EFT of completely broken $SU(2)$ symmetry in \eqref{eq_cov_dv}.

It often happens that, in presence of broken spacetime symmetries, some of the Goldstones can be algebraically eliminated in favor of the others. This is done imposing the so-called inverse Higgs constraints~\cite{IvanovIHC}. In this case, we can eliminate the Goldstones associated to the boost generators by imposing\footnote{We use that, in our convention, the boost matrix can be written as~\cite{MoninWheel}
\begin{align}
\Lambda^0_{\;\;0}=\gamma\,\quad \Lambda^0_{\;\;i}=\gamma \beta_i\,\quad \Lambda^i_{\;\;0}=\gamma\beta^i\,\quad\Lambda^i_{\;\;j}=\delta^i_{\;\;j}+(\gamma-1)\frac{\beta^i\beta_j}{\beta^2}\,, \notag
\end{align}
with the velocity related to the Goldstone by $\beta_i=\frac{\eta_i}{\eta}\tanh \eta$ and $\gamma^{2} = \frac1{1 - \beta^2}$.}
\begin{align}\label{eqIHCframons}
\nabla_i\pi_3=0\quad\implies \quad
\frac{\eta^i}{\eta}\tanh{\eta}=-\frac{D_i\chi}{D_0\chi}=-\frac{\partial_i\pi_3}{\mu}+\dots\,.
\end{align}
Crucially, thanks to the transformation properties of the covariant derivative, this constraint is compatible with all the symmetries. Consequently it is always \emph{possible} to impose it. The physical reason is that, when the system breaks spacetime symmetries, the same physical fluctuation may be described as the action of different generators. In this case, a small fluctuation generated by a boost could be obtained from the action of $Q_3$ as well~\cite{Nicolis_More,LowIHC}, making the field $\eta_i$ redundant.

Once the condition~\eqref{eqIHCframons} has been imposed, all the remaining invariants are expressed in terms of $D_\mu\chi$ and $D_\mu\alpha$ only---i.e. the covariant derivatives of the simpler completely broken $SU(2)$ theory. Without making further calculations, we know that the most general $SU(2)$ and Lorentz invariant Lagrangian written in terms of these objects is given by Eq. \eqref{eq_GenLag}.

We can also see this explicitly by writing the invariants obtained combining \eqref{eq:MCders} and \eqref{eqCovariantCCWZderivative}. To this aim, it is convenient to notice that Eq. \eqref{eqIHCframons} implies
\begin{align}\label{AppNmu}
\Lambda_\mu^{\;0}=\frac{D_\mu\chi}{\sqrt{D_\mu\chi D^\mu\chi}}\equiv n_\mu\,,\qquad
\Lambda_\mu^{\;i}\Lambda_\nu^{\;i}=-\eta_{\mu\nu}+n_\mu n_\nu\equiv P_{\mu\nu}\,.
\end{align}
Here we have conveniently defined a unit four-vector $n_\mu\simeq \delta_\mu^0+\ldots$ in the direction of the superfluid velocity and a projector $P_{\mu\nu}$ orthogonal to it. Using these quantities, the leading order invariants take the form:
\begin{align}
\begin{aligned}[c]
\nabla_0\pi_3 &=n^\mu D_\mu\chi-\mu\;,\\
 \nabla_i\alpha\nabla_i\alpha^*&=D_\mu\alpha P^{\mu\nu} D_\nu\alpha^*\;,
\end{aligned}
\qquad\qquad
\begin{aligned}[c]
\nabla_0\alpha &=n^\mu D_\mu\alpha\;,\\
\nabla_i\alpha\nabla_i\alpha &=D_\mu\alpha P^{\mu\nu} D_\nu\alpha\;.
\end{aligned}
\end{align}
The first three expressions here agree with Eq. \eqref{eq_invariants} when written in terms of the fields in \eqref{eq:coset2} using \eqref{eq:CD}. Higher order invariants are similarly obtained, for instance:
\begin{align} \notag
&\nabla_i\eta^i=\pd_\mu n^\mu\;,\\ \notag
&\nabla_0^H\nabla_0\pi_3=n^\mu\pd_\mu(n^\rho D_\rho\chi)\;,\\ \notag
&\nabla_i\alpha^*\nabla_0^H\nabla_i\alpha=-D_\mu\alpha^*P^{\mu\sigma}n^\rho
\pd_\rho(P_{\sigma\nu}D^\nu\alpha)\;,
\\ \notag
&\nabla_j\eta^i\nabla_j\eta^i=-P^{\mu\nu}\pd_\mu n^\rho\pd_\nu n_\rho\;,\\  \notag
&\nabla_0\eta^i\nabla_0\eta^i=-(n^\mu\pd_\mu n^\rho)\eta_{\rho\sigma}(n^\nu\pd_\nu n^\sigma)\;,
\\ 
&(\nabla_j\nabla_i\alpha^*)(\nabla_j\nabla_i\alpha)=-
P^{\rho\sigma}\pd_\sigma\left(P^{\mu\nu}D_\nu\alpha^*\right)
\pd_\rho\left(P_{\mu\lambda}D^\lambda\alpha\right)\;.
\end{align}
We checked up to fourth order in derivatives that all invariants obtained combining \eqref{eq:MCders} and \eqref{eqCovariantCCWZderivative} can be written contracting in a Lorentz invariant way $\pd_\mu$, $D_\mu\chi$ and $D_\mu\alpha$, as in Eq. \eqref{eq_GenLag}.

\subsection{The inverse Higgs constraint in the NREFT}\label{sec_IHC}

{
Within the spacetime coset construction presented in the previous section, there exists also the possibility of imposing an extra Inverse-Higgs constraint of the form\footnote{Of course, one could alternatively consider $\nabla_0\alpha_2=\text{Im}[\nabla_0\alpha]\simeq\dot{\alpha}_2-\mu\alpha_1=0$.} $\nabla_0\alpha_1=\text{Re}[\nabla_0\alpha]\simeq\dot{\alpha}_1+\mu\alpha_2=0$, which eliminates one of the two real components of $\alpha=\alpha_1+i\alpha_2$. Here we discuss the interpretation of this constraint within the NREFT.

In section \ref{SecNREFTLag} we showed that the NREFT describes two modes, corresponding to the gapless and the gapped Goldstones. In particular, the complex field $\pi=e^{i\chi}\alpha$ interpolates a single degree of freedom, as typical of a nonrelativistic field. However, there exists an analogous description in terms of a real field. To see this, let us rewrite the quadratic action \eqref{eqNREFTQuadraticL} to leading order in derivatives in terms of the real fields $\alpha_1$ and $\alpha_2$, with all time derivatives acting on the first and discarding total derivatives. One gets
\begin{align}
\mL\supset
-c^{(1)}\mu^3\left[\alpha_2\dot{\alpha}_1+\mu\frac{\alpha_1^2+\alpha_2^2}{2}\right]
-c_3^{(2)}\mu^2\big[(\bm \nabla \alpha_1)^2+(\bm \nabla\alpha_2)^2\big]\,.
\end{align}
Since there is no time derivative acting on it, $\alpha_2$ is an auxiliary field, which can be integrated out on its equation of motion. This gives
\begin{align}\label{eqNREFTIHC}
0=\dot{\alpha}_1+\mu\alpha_2+\mO\left(\bm \nabla^2/\mu\right)
\simeq \nabla_0\alpha_1+\mO\left(\bm \nabla^2/\mu\right)\,.
\end{align}
We hence recovered the inverse Higgs constraint\footnote{With the current parametrization the inverse Higgs constraint corresponds to the equations of motion of $\alpha_2$ only to linear order in the fields. However, the equality is true at all nonlinear orders in the Euler parametrization of the Goldstones: $\Omega=e^{i\chi Q_3}e^{i\alpha_1 Q_1}e^{i\alpha_2 Q_2}$. In other words, there is a field redefinition for which to impose the inverse Higgs constraint corresponds to integrate out $\alpha_2$ to leading orders in derivatives but to all orders in the field expansion.} $\nabla_0\alpha_{1}=0$. Since we integrated out an auxiliary field, the number of degrees of freedom and all the other properties of the action are unaffected. Indeed, plugging back the solution of \eqref{eqNREFTIHC} in the Lagrangian we find that $\alpha_1$ becomes a real field with gap $\mu$. In practice, in a nonrelativistic setting it is easier to work with a complex field, which makes particle number conservation manifest. We did not explore the possibility of building the action using only two real fields from start, e.g. working with an $SU(2)/U(1)$ coset $\Omega=e^{i\chi Q_3}e^{i\alpha_1 Q_1}$ around the background $\chi=\mu t$, $\alpha_1=0$.

\vspace{1em}

This inverse Higgs constraint was also discussed in \cite{Nicolis_More}. However, the authors there focused on a different setup, where the derivative expansion is controlled by a scale $\Lambda\gg \mu$. In that case, imposing or not the inverse Higgs constraint leads to physically distinct theories, providing a different interpretation for it. Let us briefly review these previous findings, in order to compare them with our construction.

When the inverse Higgs constraint is imposed, the construction of \cite{Nicolis_More} leads to an EFT describing the gapless and the gapped Goldstone, with cutoff $\Lambda\gg\mu$. In this setup, the symmetry is partially restored in the limit $\mu\rightarrow 0$, if this limit exists.\footnote{This is not obvious even for $\Lambda\gg\mu$, since the cutoff itself might depend on the chemical potential, e.g. as $\Lambda^2\sim f\mu$ with $f\gg\mu$; see \cite{Nicolis_More} for details.}
As discussed in the introduction of section \ref{SecNREFT}, this EFT applies for instance in the linear sigma model for $m^2<0$ when the radial mode is much heavier than the gapped Goldstone, i.e. when $|m^2|\gg\mu^2$. 

The situation is different when the inverse Higgs constraint is not imposed. Indeed, when $\Lambda\gg\mu$, the leading order quadratic Lagrangian for the complex field $\alpha$ is second order in time derivatives, implying that $\alpha$ interpolates two modes rather than one as in our nonrelativistic construction. One mode is the gapped Goldstone, while the mass of the other depends on the coefficients of the Lagrangian and it is formally proportional to $\mu$. This mode is usually referred to as a gapped Goldstone with unfixed gap~\cite{Nicolis_More}. In this case, if the limit $\mu\rightarrow 0$ is smooth, the theory breaks the internal $SU(2)$ symmetry completely also at zero chemical potential; the extra mode then provides the third Goldstone required by the relativistic Goldstone theorem.

In general, the presence of the unfixed gap mode and its properties are not fixed by the symmetry breaking pattern only and depend on the structure of the theory at scales $\Lambda\gg\mu$. Thus, for the purposes of our construction in which the chemical potential itself provides the cutoff, this mode, if present in the UV theory, behaves rather like any other matter field and is thus integrated out in our setup. The nonrelativistic EFT, similarly to the standard relativistic CCWZ construction, provides the minimal structure required to realize nonlinearly all the symmetries; in practice, this means that the NREFT describes only the gapless and the gapped Goldstones. Of course, while we expect this simple setup to correspond to the most generic situation, specific theories may contain additional light degrees of freedom, e.g. gauge fields, which can be added to the EFT in the standard way.}

\section{NREFT details}\label{AppNREFT}

\subsection{NREFT action to \texorpdfstring{$\mO(\pd^4)$}{Od4}}\label{AppNREFTAction}

In this section, we write the Lagrangian for the non-relativistic effective theory to fourth order in derivatives. To this aim, we find a convenient bookkeeping tool to use the invariants written using the spacetime coset construction presented in appendix \ref{sec:cosetgeneral}. We assume parity invariance for simplicity.

The effective nonrelativistic Lagrangian is written using the prescription presented in section~\ref{SecNREFTCoset}, namely imposing the $U(1)$ invariance $\pi\rightarrow e^{i\xi}\pi$ and using the nonrelativistic derivative \eqref{nonrelativisticD}. In the notation of the previous section, the latter amounts at building higher derivative terms using, rather than the one given in Eq. \eqref{eqCovariantCCWZderivative}, the following covariant derivative:
\begin{align}
e^{\;\;\mu}_a \hat{\partial}_\mu+\frac{i}{2}\omega^{ij}_aJ_{ij}=
\nabla_a^H
+i\left(\mu+\nabla_0\pi_3\right)\delta_a^0 [Q_3,\cdot]\;.
\end{align}
In practice, we performed calculations using the following 
\begin{align}
\wD_a\equiv \nabla_a^H
+i\,\mu\delta_a^0 [Q_3,\cdot]\;.
\end{align}
This definition corresponds to a slightly different form of the nonrelativistic derivative, obtained multiplying $D_\mu\chi$ in Eq. \eqref{nonrelativisticD} by $\mu/\sqrt{D_\mu\chi D^\mu\chi}$. As commented below that equation, this redefinition does not affect the key property \eqref{nonrelativisticDprop}, which is needed in order to have a well-structured derivative expansion.

We can now proceed to formally write the Lagrangian in a $\nabla/\mu$ expansion as
\begin{align}\label{eqAppLag4}
\mL=\mL^{(1)}_{\nabla}+\mL^{(2)}_{\nabla}+\mL^{(3)}_{\nabla}+\mL^{(4)}_{\nabla}+\ldots\,,
\end{align}
where $\mL^{(i)}_{\nabla}$ contains all terms which are of order $i$ in terms of $\nabla$'s covariant derivatives. We have:
\begin{align}
\mL_\nabla^{(1)}/\mu^3&=c^{(1)}\nabla_0\pi_3\,,\\
\mL_\nabla^{(2)}/\mu^2&=c^{(2)}_1(\nabla_0\pi_3)^2
+c^{(2)}_2|\nabla_0\alpha|^2-c^{(2)}_3|\nabla_i\alpha|^2\,,\\ 
\mL_\nabla^{(3)}/\mu&=c^{(3)}_1(\nabla_0\pi_3)^3
+c^{(3)}_2 \nabla_0\pi_3|\nabla_0\alpha|^2
+c^{(3)}_3\nabla_0\pi_3|\nabla_i\alpha|^2
\notag \\& 
+c_4^{(3)}\left[i\nabla_0\alpha^*\wD_0\left(\nabla_0\alpha\right)+c.c.\right]
+c_5^{(3)}\left[i\nabla_i\alpha^*\wD_0\left(\nabla_i\alpha\right)+c.c.\right]
\notag \\& 
+c_{6}^{(3)}\left[\nabla_i\alpha^*\wD_i\left(\nabla_0\alpha\right)+c.c.\right]
+c_{7}^{(3)}\left[i\nabla_i\alpha^*\wD_i\left(\nabla_0\alpha\right)+c.c.\right]
\notag \\&
+c_{8}^{(3)}\nabla_0\pi_3(\mu\nabla_i\eta^i)\,. \label{eqNREFTLagrangiand3} 
\end{align}
We can expand these in terms of the $SU(2)$ covariant derivatives in Eq. \eqref{eq_cov_dv} and their derivatives. Doing so and defining $D_\mu\pi_3\equiv D_\mu\chi-\mu\delta_\mu^0$, we can rewrite the Lagrangian in a standard derivative expansion:
\begin{align}
\mL^{(1)}/\mu^3&=c^{(1)}D_0\pi_3\,,\\
\mL^{(2)}/\mu^2&=c^{(2)}_1(D_0\pi_3)^2-\frac{c^{(1)}}{2}(D_i\pi_3)^2
+c^{(2)}_2|D_0\alpha|^2-c^{(2)}_3|D_i\alpha|^2
\,,\\ \nonumber
\mL^{(3)}/\mu&=
\left[\frac{c^{(1)}}{2}-c^{(2)}_1\right]D_0\pi_3 (D_i\pi_3)^2
+\left[c^{(2)}_2-c^{(2)}_3-c_{7}^{(3)}\right]\left(D_0\alpha^* D_i\alpha D_i\pi_3+\text{c.c.}\right)
\\& \nonumber
+c_{6}^{(3)}\left(iD_i\alpha^* D_i\pi_3D_0\alpha +\text{c.c.}\right)
+c^{(3)}_1(D_0\pi_3)^3
+c^{(3)}_2 D_0\pi_3|D_0\alpha|^2
+c^{(3)}_3 D_0\pi_3|D_i\alpha|^2
\\& 
+c_4^{(3)}\left[iD_0\alpha^*(\pd_0+i\mu)\left(D_0\alpha\right)+\text{c.c.}\right]
+c_5^{(3)}\left[iD_i\alpha^*(\pd_0+i\mu)\left(D_i\alpha\right)+\text{c.c.}\right]
\\& 
+c_{6}^{(3)}\left[D_i\alpha^*\pd_i\left(D_0\alpha\right)+\text{c.c.}\right]
+c_{7}^{(3)}\left[iD_i\alpha^*\pd_i\left(D_0\alpha\right)+\text{c.c.}\right]
-c_{8}^{(3)}D_0\pi_3(\pd_i D_i\pi_3)\,.\notag
\end{align}
Notice that terms with $D_i\pi_3$ always appear from the expansion of the $\nabla$ covariant derivatives in connection with lower derivative ones. \\
The fourth order in derivatives can be constructed similarly. Here we just report the fourth order term in \eqref{eqAppLag4}
\begin{align} \notag
\mL^{(4)}_\nabla&=c^{(4)}_1(\nabla_0\pi_3)^4
+c^{(4)}_2 (\nabla_0\pi_3)^2|\nabla_0\alpha|^2
+c^{(4)}_3(\nabla_0\pi_3)^2|\nabla_i\alpha|^2 \\&
+c_4^{(4)}|\nabla_0\alpha|^4 
+c_5^{(4)}|\nabla_0\alpha|^2|\nabla_i\alpha|^2 \notag
+c^{(4)}_6\left[(\nabla_i\alpha)^2(\nabla_0\alpha^*)^2+\text{c.c.}\right]
\\&
+c^{(4)}_7\left[i(\nabla_i\alpha)^2(\nabla_0\alpha^*)^2+\text{c.c.}\right]
+c^{(4)}_8|\nabla_i\alpha|^2|\nabla_j\alpha|^2 \notag \\
&+c^{(4)}_9(\nabla_i\alpha)^2(\nabla_j\alpha^*)^2
+c^{(4)}_{10}(\wD_0\nabla_0\pi_3)|\nabla_0\alpha|^2
+c^{(4)}_{11}(\nabla_0\pi_3)\left[i\nabla_0\alpha^*\wD_0\nabla_0\alpha+\text{c.c.}\right] \notag
\\&
+c^{(4)}_{12}(\wD_0\nabla_0\pi_3)|\nabla_i\alpha|^2
+c^{(4)}_{13}(\nabla_0\pi_3)\left[i\nabla_i\alpha^*\wD_0\nabla_i\alpha+\text{c.c.}\right] \notag
\\&
+c^{(4)}_{14}(\wD_i\nabla_0\pi_3)\left[\nabla_i\alpha^*\nabla_0\alpha+\text{c.c.}\right]
+c^{(4)}_{15}(\wD_i\nabla_0\pi_3)\left[i\nabla_i\alpha^*\nabla_0\alpha+\text{c.c.}\right] \notag
\\&
+c^{(4)}_{16}\nabla_0\pi_3\left[\wD_i(\nabla_0\alpha)\nabla_i\alpha^*+\text{c.c.}\right]
+c^{(4)}_{17}\nabla_0\pi_3\left[i\wD_i(\nabla_0\alpha)\nabla_i\alpha^*+\text{c.c.}\right] \notag
\\&
+c^{(4)}_{18}(\wD_0\nabla_0\pi_3)^2+c^{(4)}_{19}(\wD_i\nabla_0\pi_3)^2
+c^{(4)}_{20}|\wD_0\nabla_0\alpha|^2 \notag
\\&
+c^{(4)}_{21}|\wD_i\nabla_0\alpha|^2
+c^{(4)}_{22}\left[\wD_i\nabla_i\alpha^*\wD_0\nabla_0\alpha+\text{c.c.}\right]
+c^{(4)}_{23}\left[i\wD_i\nabla_i\alpha^*\wD_0\nabla_0\alpha+\text{c.c.}\right] \notag
\\&
+c^{(4)}_{24}|\wD_0\nabla_i\alpha|^2+c^{(4)}_{25}|\wD_i\nabla_i\alpha|^2
+c^{(4)}_{26}|\wD_j\nabla_i\alpha|^2+c^{(4)}_{27}\mu^2\nabla_0\eta^i\nabla_0\eta^i \notag
+c^{(4)}_{28}\mu^2(\nabla_i\eta^i)^2
\\&
+c^{(4)}_{29}\mu^2\nabla_i\eta^j\nabla_i\eta^j
+c^{(4)}_{30}\mu\nabla_i\eta^i\wD_0\nabla_0\pi_3
+c^{(4)}_{31}\mu(\nabla_0\pi_3)^2\nabla_i\eta^i
+c^{(4)}_{32}\mu|\nabla_0\alpha|^2\nabla_i\eta^i \notag
\\&
+c^{(4)}_{33}\mu|\nabla_i\alpha|^2\nabla_j\eta^j
+c^{(4)}_{34}\mu\nabla_i\eta^j\left[\nabla_i\alpha\nabla_j\alpha^*+\text{c.c.}\right]
+c^{(4)}_{35}\mu\nabla_i\eta^j\left[i\nabla_i\alpha\nabla_j\alpha^*+\text{c.c.}\right] \notag
\\&
+c^{(4)}_{36}\mu\nabla_0\eta^i\left[\nabla_i\alpha\nabla_0\alpha^*+\text{c.c.}\right]
+c^{(4)}_{37}\mu\nabla_0\eta^i\left[i\nabla_i\alpha\nabla_0\alpha^*+\text{c.c.}\right]\,. 
\end{align}
We did not write terms which effectively contribute at fifth order in derivatives 
after expanding the $\nabla$'s as before. 

\subsection{Feynman rules to leading order in \texorpdfstring{$\pd/\mu$}{d/mu}}\label{AppNREFTFeynamRules}

Before introducing a process dependent velocity power counting, it might be useful to consider a power counting in $\pd/\mu$. Here we list the Feynman rules to leading order within this counting. We use the field parametrization \eqref{eq:coset2}. Black solid lines correspond to gapped Goldstones with four-momentum $p=(\mu+\epsilon,\bm{p})$, while dashes stand for gapless Goldstones, whose four-momentum is denoted as $k=(\omega,\bm{k})$.
\begin{itemize}
\item \textbf{$|\pi|^2\pi_3$ vertex:} 
\vspace*{0.4cm}
\begin{align}
\includegraphics*[scale=1.1]{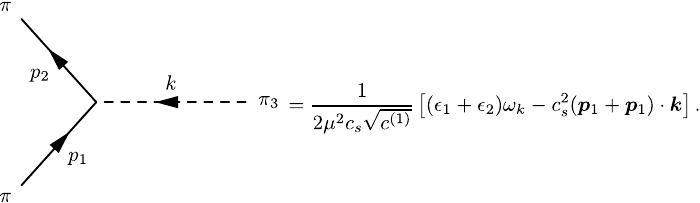} 
\end{align}
\item \textbf{$\pi^3_3$ vertex:} 
\vspace*{0.4cm}
\begin{align}
\includegraphics*[scale=1.1]{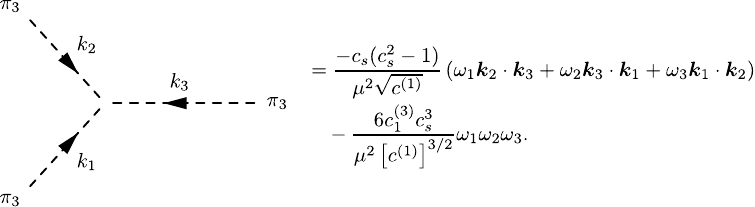} 
\end{align}
\item \textbf{$|\pi|^4$ vertex:} 
\vspace*{0.4cm}
\begin{align}
\includegraphics*[scale=1.1]{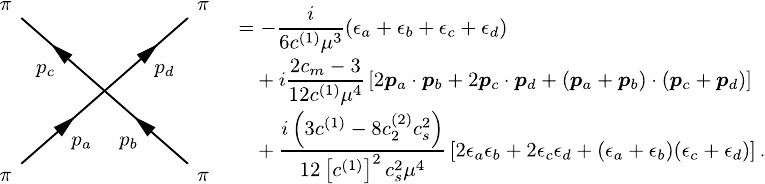} 
\end{align}
\item \textbf{$|\pi|^2\pi_3^2$ vertex:} 
\vspace*{0.4cm}
\begin{align}
\includegraphics*[scale=1.1]{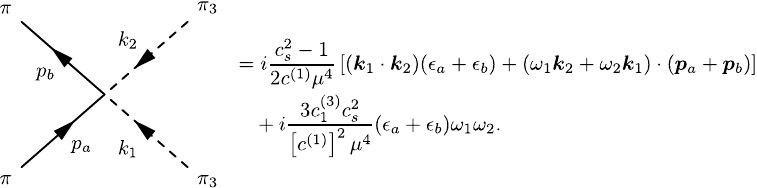} 
\end{align}
\item \textbf{$\pi_3^4$ vertex:} 
\vspace*{0.4cm}
\begin{align}
\includegraphics*[scale=1.08]{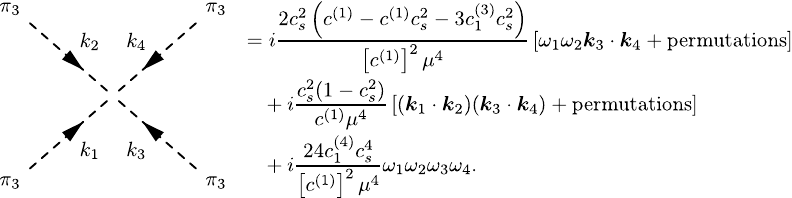} 
\end{align}
\end{itemize}

\subsection{Coefficients of \texorpdfstring{$\pi\pi$}{pipi} scattering to order \texorpdfstring{$\mO(v^4)$}{Ov4}}\label{AppNREFTscattering}

The coefficients of the $\pi$ dispersion relation \eqref{eqNREFTdispersionGappedNGB2} to subleading order is given by
\begin{align}
c_m^{(2)}=16 \frac{ c^{(1)} \left[c^{(1)} \left(c^{(4)}_{25}+c^{(4)}_{26}\right)+4 c^{(2)}_3 \left(c^{(3)}_5+c^{(3)}_7\right)\right]+4 c^{(2)}_2 \left(c^{(2)}_3\right)^2}{(c^{(1)})^3}\,.
\end{align}
The $b_i$'s in \eqref{eqNREFTscattering2} are
\begin{align}\label{eqNREFTappB1}
b_1&=-\frac{c^{(1)} c_m^2 (c_s-1)-4 c_s \left[c_m (c^{(2)}_2+c^{(3)}_3)+2 c^{(3)}_5 (c_m-1)-2 c^{(3)}_7\right]}{8 c_s}\,,\\ \label{eqNREFTappB2}
b_2 &=\frac{1}{4} \left(c_m 
\left[4 c_{3}^{(3)}-3 c^{(1)} c_m+4 c_{2}^{(2)} (2 c_m+3)\right]
-2 c^{(1)} c_m^{(2)}\right) \notag \\
&\quad+6 c_{5}^{(3)} (c_m-1)+c_{7}^{(3)} (4 c_m-6)\,,\\
b_3 &=-4 \left(2 c_{5}^{(3)}-4 c_{9}^{(4)}+2 c_{26}^{(4)}+c_{29}^{(4)}+2 c_{35}^{(4)}\right)\,,\\
b_4&=2 \left(2 c_{5}^{(3)}+4 c_{8}^{(4)}+2 c_{26}^{(4)}+c_{29}^{(4)}+2 c_{35}^{(4)}\right)\,.
\end{align}

\subsection{Loops in dimensional regularization}\label{SecNREFTLoops}

We can regulate the NREFT at quantum level with a hard space cutoff $\Lambda\lesssim\mu$. However powers of the cutoff spoil power counting \cite{KaplanEFT} and complicate computations. It is hence preferable to use a mass independent regulator, such as dimensional regularization. In a nonrelativistic EFT, if this is done na\"ively retaining the standard form of propagators, loops involving both massive and massless particle become dominated by hard momenta $|\bm k |\sim\mu$, which should not enter in the NREFT computations~(c.f~\cite{HoangNRQCD} within the context of NRQCD). This is due to the fact that the gapped dispersion relation $k_0\sim\bm k^2 /\mu$ and the gapless one $k_0\sim |\bm k |$ can be simultaneously satisfied only for $|\bm k |\sim\mu$. A consistent formulation of NREFTs with both heavy and light fields was devised by Griesshammer \cite{Gri1,Gri2}, as a development of the method of regions~\cite{Beneke}, and then further refined with the formulation of vNRQCD \cite{vNRQCD,Manohar_ZeroBin}. In this appendix we review the key points and their application to our EFT, focusing on the power counting of diagrams. We refer to the original works for details.

The first step is to identify a consistent set of modes, according to their scaling with velocity $v$. According to standard NRQCD results \cite{HoangNRQCD,RothTasi}, these are given by soft, potential and ultrasoft modes listed in \eqref{eqSoftUltraSoftPotential}. Fields are split accordingly as explained in section \ref{SecNREFTtreelevel}.
To enforce power counting, one should retain in the denominators of propagators only momenta with the same scaling in $v$, expanding the subleading ones in an infinite series. 
In particular, they will be given by \footnote{Naively performing these expansions inside loops sometimes leads to unphysical \emph{pinch} singularities, e.g. in box integrals. However, a careful analysis shows that these arise from an over-counting of the contribution of a certain region and that loops are indeed regular after the proper \emph{zero-bin} subtractions have been performed \cite{Manohar_ZeroBin}. These subtleties do not affect the simple power counting rules that we discuss here, hence we will neglect them in what follows.}
\begin{align}\label{eqPropagators}
\begin{split}
G_{\pi_3}^\text{s}(\omega,\bm k)=G_{\pi_3}^\text{us}(\omega,\bm k)=\frac{i}{\omega^2-c_s^2\bm k^2}\,,\qquad & G_{\pi_3}^\text{p}(\omega,\bm k)=\frac{-i}{c_s^2\bm k^2}\sum_{n=0}^\infty\left(\frac{\omega^2}{c_s^2\bm k^2}\right)^n\,, \\
G_\pi^\text{s}(\epsilon,\bm p)=G_\pi^\text{us}(\epsilon,\bm p)=\frac{i}{\epsilon}\sum_{n=0}^\infty\left(\frac{c_m\bm p^2}{2\mu \epsilon}\right)^n\,,& \qquad G_\pi^\text{p}(\epsilon,\bm p)=\frac{i}{\epsilon-\frac{c_m\bm p^2}{2\mu}}\,,
\end{split}
\end{align}
where we omitted the $+i 0$ prescription. 
For instance, the soft $G_{\pi}^\text{s}(\epsilon,\bm p)$ propagator and the potential $G_{\pi}^\text{p}(\epsilon,\bm p)$ propagators are not equivalent beyond tree-level, since infinite sums and integration do not commute in dimensional regularization \cite{Beneke}.
After the splitting into different modes is performed, and hence all propagators are properly expanded, all loops in dimensional regularization are made only of light scales. This also makes it straightforward to power count diagrams in $v$.

As a simple illustration, consider the one-loop correction to the $G_{\pi}^p(\epsilon,\bm p)$ propagator\footnote{We neglected a scaleless tadpole vanishing in dimensional regularization.}
\begin{align}
\includegraphics*[scale=1.2]{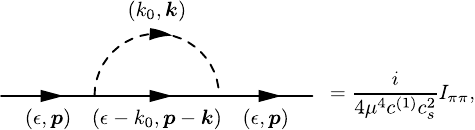} 
\end{align}
where in a hard cutoff approach we would write the loop integral as
\begin{align}\label{eqPiPiPropLoopNaive}
I_{\pi\pi}=-i\int \frac{d^4k}{(2\pi)^4}
\frac{\left[(2\epsilon-k_0)k_0-c_s^2(2\vp-\vk)\cdot\vk\right]^2}{
\left[(\epsilon-k_0)-c_m\frac{(\vp-\vk)^2}{2\mu}+i0\right]
\left(k_0^2-c_s^2\vkk+i0\right)}.
\end{align}
To perform this computation in $d=4-\varepsilon$ dimensions, we need to take into account four different integrals, depending on the specific modes running in the loop:
\begin{enumerate}
\item $\pi_3^s:(k_0,\vk)\sim (\mu v,\mu v)$ and 
$\pi^s:(\epsilon-k_0,\vp-\vk)\sim(\mu v,\mu v)$;
\item $\pi_3^p:(k_0,\vk)\sim  (\mu v^2,\mu v)$ and 
$\pi^p:(\epsilon-k_0,\vp-\vk)\sim(\mu v^2,\mu v)$;
\item $\pi_3^p:(k_0,\vk)\sim  (\mu v^2,\mu v)$ and 
$\pi^{us}:(\epsilon-k_0,\vp-\vk)\sim(\mu v^2,\mu v^2)$;
\item $\pi_3^{us}:(k_0,\vk)\sim  (\mu v^2,\mu v^2)$ and 
$\pi^p:(\epsilon-k_0,\vp-\vk)\sim(\mu v^2,\mu v)$.
\end{enumerate}
Consider for illustration the $\pi^\text{s}-\pi_3^\text{s}$ loop. We have $k_0\gg\epsilon,(\vp-\vk)^2/\mu$, hence we should enforce this expanding the gapped Goldstone propagator in an infinite series
\begin{align}
\frac{i}{(\epsilon-k_0)-c_m\frac{(\vp-\vk)^2}{2\mu}+i0}\quad
\longrightarrow\quad
\frac{i}{-k_0+i0}\left[1+
\frac{\epsilon-c_m\frac{(\vp-\vk)^2}{2\mu}}{k_0-i0}+\ldots
\right]
\end{align}
The integral here is:
\begin{align}
I_{\pi\pi}^{(1)}=-iM^\varepsilon
\int \frac{d^dk}{(2\pi)^d}
\frac{\left[\ldots\right]^2}{
\left(-k_0+i0\right)
\left(k_0^2-c_s^2\vkk+i0\right)}\left[
1+\frac{\epsilon-c_m\frac{(\vp-\vk)^2}{2\mu}}{k_0-i0}+\ldots
\right]=0,
\end{align}
where $M$ is the sliding scale. The loop vanishes since, after performing the $k^0$ integration with the residue's theorem, the integral can be divided in a sum of contributions proportional to $\int d^{d-1}{\bm k}/|{\bf k}|^n=0$.
Similarly one can check that the $\pi_3^\text{p}-\pi^\text{p}$ and $\pi_3^\text{p}-\pi^\text{s}$ loops vanish\footnote{Within this approach this is a common fact, for instance one can prove that $\pi^\text{us}$ never contributes inside loops \cite{Gri1}.} to all orders in $v$.\\
The only nontrivial contribution comes from the ultrasoft $\pi_3^\text{us}-\pi^\text{p}$ loop.
We have $\vkk\ll\vpp$, implying that the $\pi^\text{p}$ propagator should be expanded as
\begin{align}\label{eqPotentialLoopPiPropagator2}
\frac{i}{(\epsilon-k_0)-c_m\frac{(\vp-\vk)^2}{2\mu}+i0}\quad
\longrightarrow\quad
\frac{i}{(\epsilon-k_0)-c_m\frac{\vpp}{2\mu}+i0}
\left(1-\frac{c_m\frac{\vp\cdot\vk}{\mu}}{
\epsilon-k_0
-c_m\frac{\vpp}{2\mu}}+\ldots
\right).
\end{align}
We can power count the measure according to the momentum of the softest propagator, which sets the size of the \emph{integration box}. 
In this case thus $d^4k\sim\mu^4v^8$. The leading contribution is
\begin{align}\label{IpipiExamplePowerCounting}
I_{\pi\pi}^{(4)}=
M^{\varepsilon}
\int \frac{d^dk}{(2\pi)^d}
\frac{-ic_s^4(\vp\cdot\vk)^2}{\left[\left(\epsilon-k_0\right)
-c_m\frac{\vpp}{2\mu}+i0\right]
\left(k_0^2-
c_s^2\vkk+i0\right)}
\sim \mO\left(v^8\right).
\end{align}
The integral is simple to perform, giving an $\mO(v^8)$ contribution:
\begin{align}\label{IpipiExampleResult}
I_{\pi\pi}^{(4)}= \frac{\vpp\left(\epsilon-c_m\frac{\vpp}{2\mu}\right)^3}{3  \pi ^2 c_s}
\left\{\frac{1}{\varepsilon }-
\log \left(\frac{\epsilon-c_m\frac{\vpp}{2\mu}+i0 }{-c_s M}\right)-\frac{\gamma }{2}+\frac{4}{3}+\frac{\log \pi }{2}
\right\}.
\end{align}
The divergence renormalizes the Lagrangian term 
$\frac{1}{\mu^5}\nabla_i\pi^*\left(i\wD_0-\frac{c_m}{2\mu}\wD_i\wD_i\right)^3\nabla_i\pi$ $=\frac{1}{\mu^5}D^\lambda\pi^* P_{\lambda\mu}\left\{
\left[i n^\rho\hd_\rho -\frac{c_m}{2\mu}\hd_\rho\left(P^{\rho\sigma}\hd_\sigma\right)\right]^3P^{\mu\nu}D_\nu\pi\right\}
$, in the notation of appendix \ref{sec:cosetgeneral}. In practice many tree-level higher derivative terms have to be taken into account at the lower orders.

\begin{figure}[t]
\centering
\includegraphics[scale=1.1]{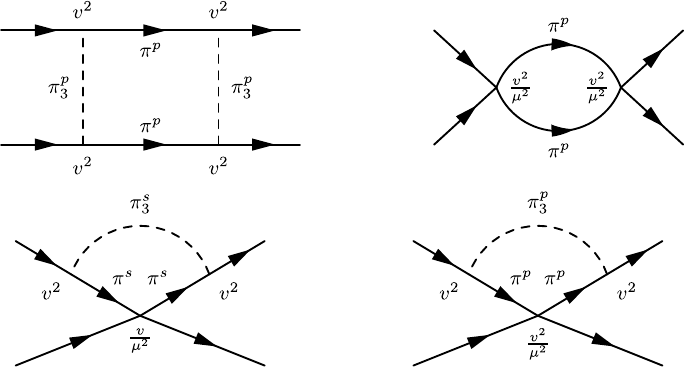}
\caption{Leading loop topologies which correct the contact interaction in $\pi\pi$ scattering. The scaling of vertices and the modes in the loop are displayed.}
\label{FigPiPiLoops}
\end{figure}

In \eqref{IpipiExampleResult} we found a $\sim
\log\left(\mu v^2/M\right)$ contribution. Indeed in general ultrasoft loops give rise to logarithms of the \emph{ultrasoft} scale $\mu v^2$. Instead soft and potential loops lead to logarithms of the $\emph{soft}$ scale $\mu v$ \cite{vNRQCD}. For instance, the leading loop contribution to $\pi^\text{p}_3\pi^\text{p}_3$ potential propagator comes from a soft loop and takes the form
\begin{align}\label{eqPi3Pi3PropagatorLoop}
\includegraphics*[scale=1.2]{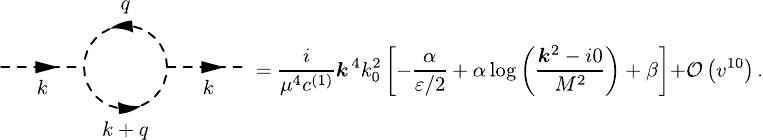} 
\end{align}
Finally, one can now consistently power count loop contributions to the $\pi\pi$ elastic scattering, computed to $\mO(v^4)$ at tree-level in Sec. \ref{SecNREFTtreelevel}. Using the Feynman rules in \ref{AppNREFTFeynamRules}, one easily concludes that the first corrections arise only at $\mO(v^5)$. Specifically, three kinds of loop corrections exist. First, corrections to the $G_{\pi_3}^\text{p}$ propagator in exchange diagrams of fig. \ref{FigPi3Piscattering}, which however start at $\mO(v^8)$ as Eq. \eqref{eqPi3Pi3PropagatorLoop} shows. Then corrections to the $\pi_3^\text{p}|\pi^\text{p}|^2$ vertex appearing in the same kind of diagrams. For instance, the leading correction in this class is given by a loop of $\pi^\text{p}$ and $\pi_3^\text{p}$:
\begin{align}
\includegraphics*[scale=1.2]{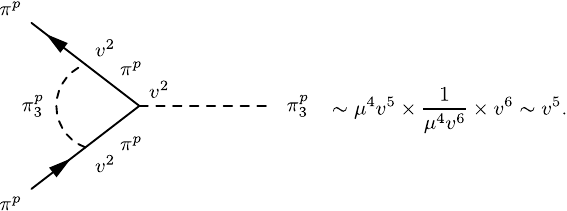} 
\end{align}
Here we showed explicitly the scaling of the vertices with $v$ and we power counted the result as \emph{measure}$\times$\emph{propagators}$\times$\emph{vertices}.
Finally we have those that we can interpret as corrections to the contact vertex in fig. \ref{FigPiPiscattering}. The leading corrections in this class are also $\mO(v^5)$ and are displayed in figure \ref{FigPiPiLoops}.

We remark that this formulation of the NREFT differs in some points from the modern vNRQCD \cite{vNRQCD}. First, we did not separate explicitly the fields momenta in soft, potential and ultra-soft components in the Lagrangian, as it is customarily done in NRQCD \cite{RothTasi}. Of course, this is possible and it might be useful in performing more refined computations, especially to account for the proper \emph{zero-bin} subtractions \cite{Manohar_ZeroBin}.
Furthermore, here off-shell modes are not integrated out explicitly and the \emph{pull-up} mechanism is not explicitly implemented \cite{vNRQCD}, i.e. we do not renormalize soft and ultrasoft fields separately. 
These differences stem from the fact that we want to preserve the nonlinearly realized $SU(2)$ invariance in the Lagrangian, which relates the different modes of $\pi_3$ and $\pi$. In particular, this implies that all modes of a given operator have the same anomalous dimensions \cite{Gri2}, differently than in vNRQCD. There, renormalizing them separately allows to efficiently resum logarithms of both the \emph{soft} and \emph{ultrasoft} scale, via the velocity Renormalization Group.\footnote{For instance, in vNRQCD ultrasoft and soft fields are allowed to interact via gauge couplings evaluated at different sliding scales.} This is not possible within our approach, but it is only a minor drawback. Indeed, as typical for Goldstone bosons, all interactions are irrelevant, so that logarithms are always multiplied by powers of the velocity.

\bibliography{Biblio}

\providecommand{\href}[2]{#2}\begingroup\raggedright\begin{thebibliography}{10}

\bibitem{Goldstone}
J.~Goldstone, A.~Salam and S.~Weinberg, \emph{{Broken Symmetries}},
  \href{https://doi.org/10.1103/PhysRev.127.965}{\emph{Phys. Rev.} {\bfseries
  127} (1962) 965}.

\bibitem{Nambu}
Y.~Nambu, \emph{Quasi-particles and gauge invariance in the theory of
  superconductivity},
  \href{https://doi.org/10.1103/PhysRev.117.648}{\emph{Phys. Rev.} {\bfseries
  117} (1960) 648}.

\bibitem{CCWZ1}
S.~R. Coleman, J.~Wess and B.~Zumino, \emph{{Structure of phenomenological
  Lagrangians. 1.}},
  \href{https://doi.org/10.1103/PhysRev.177.2239}{\emph{Phys. Rev.} {\bfseries
  177} (1969) 2239}.

\bibitem{CCWZ2}
C.~G. Callan, Jr., S.~R. Coleman, J.~Wess and B.~Zumino, \emph{{Structure of
  phenomenological Lagrangians. 2.}},
  \href{https://doi.org/10.1103/PhysRev.177.2247}{\emph{Phys. Rev.} {\bfseries
  177} (1969) 2247}.

\bibitem{Weinberg2}
S.~Weinberg, \emph{{The quantum theory of fields. Vol. 2: Modern
  applications}}. Cambridge University Press, 2013.

\bibitem{LangeNRGoldstones}
R.~V. Lange, \emph{{Goldstone Theorem in Nonrelativistic Theories}},
  \href{https://doi.org/10.1103/PhysRevLett.14.3}{\emph{Phys. Rev. Lett.}
  {\bfseries 14} (1965) 3}.

\bibitem{NielsenNRGoldstones}
H.~B. Nielsen and S.~Chadha, \emph{{On How to Count Goldstone Bosons}},
  \href{https://doi.org/10.1016/0550-3213(76)90025-0}{\emph{Nucl. Phys.}
  {\bfseries B105} (1976) 445}.

\bibitem{WatanabeRedundancies}
H.~Watanabe and H.~Murayama, \emph{{Redundancies in Nambu-Goldstone Bosons}},
  \href{https://doi.org/10.1103/PhysRevLett.110.181601}{\emph{Phys. Rev. Lett.}
  {\bfseries 110} (2013) 181601}
  [\href{https://arxiv.org/abs/1302.4800}{{\ttfamily 1302.4800}}].

\bibitem{BraunerNonRelNGB}
T.~Brauner, \emph{{Spontaneous Symmetry Breaking and Nambu-Goldstone Bosons in
  Quantum Many-Body Systems}},
  \href{https://doi.org/10.3390/sym2020609}{\emph{Symmetry} {\bfseries 2}
  (2010) 609} [\href{https://arxiv.org/abs/1001.5212}{{\ttfamily 1001.5212}}].

\bibitem{Morchio:1987aw}
G.~Morchio and F.~Strocchi, \emph{{Effective Non-Symmetric Hamiltonians and
  Goldstone Boson Spectrum}},
  \href{https://doi.org/10.1016/0003-4916(88)90046-2}{\emph{Annals Phys.}
  {\bfseries 185} (1988) 241}.

\bibitem{Strocchi:2008gsa}
F.~Strocchi, \emph{{Symmetry Breaking}}, vol.~732. 2008,
  \href{https://doi.org/10.1007/978-3-540-73593-9}{10.1007/978-3-540-73593-9}.

\bibitem{Nicolis_Theorem}
A.~Nicolis and F.~Piazza, \emph{{Implications of Relativity on Nonrelativistic
  Goldstone Theorems: Gapped Excitations at Finite Charge Density}},
  \href{https://doi.org/10.1103/PhysRevLett.110.011602,
  10.1103/PhysRevLett.110.039901}{\emph{Phys. Rev. Lett.} {\bfseries 110}
  (2013) 011602} [\href{https://arxiv.org/abs/1204.1570}{{\ttfamily
  1204.1570}}].

\bibitem{Nicolis_More}
A.~Nicolis, R.~Penco, F.~Piazza and R.~A. Rosen, \emph{{More on gapped
  Goldstones at finite density: More gapped Goldstones}},
  \href{https://doi.org/10.1007/JHEP11(2013)055}{\emph{JHEP} {\bfseries 11}
  (2013) 055} [\href{https://arxiv.org/abs/1306.1240}{{\ttfamily 1306.1240}}].

\bibitem{WatanabeMNGB}
H.~Watanabe, T.~Brauner and H.~Murayama, \emph{{Massive Nambu-Goldstone
  Bosons}}, \href{https://doi.org/10.1103/PhysRevLett.111.021601}{\emph{Phys.
  Rev. Lett.} {\bfseries 111} (2013) 021601}
  [\href{https://arxiv.org/abs/1303.1527}{{\ttfamily 1303.1527}}].

\bibitem{Nicolis_SSP}
A.~Nicolis and F.~Piazza, \emph{{Spontaneous Symmetry Probing}},
  \href{https://doi.org/10.1007/JHEP06(2012)025}{\emph{JHEP} {\bfseries 06}
  (2012) 025} [\href{https://arxiv.org/abs/1112.5174}{{\ttfamily 1112.5174}}].

\bibitem{Nicolis_Zoology}
A.~Nicolis, R.~Penco, F.~Piazza and R.~Rattazzi, \emph{{Zoology of condensed
  matter: Framids, ordinary stuff, extra-ordinary stuff}},
  \href{https://doi.org/10.1007/JHEP06(2015)155}{\emph{JHEP} {\bfseries 06}
  (2015) 155} [\href{https://arxiv.org/abs/1501.03845}{{\ttfamily
  1501.03845}}].

\bibitem{Maris:1977zz}
H.~J. Maris, \emph{{Phonon-phonon interactions in liquid helium}},
  \href{https://doi.org/10.1103/RevModPhys.49.341}{\emph{Rev. Mod. Phys.}
  {\bfseries 49} (1977) 341}.

\bibitem{Brauner}
T.~Brauner and M.~F. Jakobsen, \emph{{Scattering amplitudes of massive
  Nambu-Goldstone bosons}},
  \href{https://doi.org/10.1103/PhysRevD.97.025021}{\emph{Phys. Rev.}
  {\bfseries D97} (2018) 025021}
  [\href{https://arxiv.org/abs/1709.01251}{{\ttfamily 1709.01251}}].

\bibitem{MoninCFT}
A.~Monin, D.~Pirtskhalava, R.~Rattazzi and F.~K. Seibold, \emph{{Semiclassics,
  Goldstone Bosons and CFT data}},
  \href{https://doi.org/10.1007/JHEP06(2017)011}{\emph{JHEP} {\bfseries 06}
  (2017) 011} [\href{https://arxiv.org/abs/1611.02912}{{\ttfamily
  1611.02912}}].

\bibitem{NRQED}
W.~E. Caswell and G.~P. Lepage, \emph{{Effective Lagrangians for Bound State
  Problems in QED, QCD, and Other Field Theories}},
  \href{https://doi.org/10.1016/0370-2693(86)91297-9}{\emph{Phys. Lett.}
  {\bfseries 167B} (1986) 437}.

\bibitem{Labelle}
P.~Labelle, G.~P. Lepage and U.~Magnea, \emph{{Order m-alpha**8 contributions
  to the decay rate of orthopositronium}},
  \href{https://doi.org/10.1103/PhysRevLett.72.2006}{\emph{Phys. Rev. Lett.}
  {\bfseries 72} (1994) 2006}
  [\href{https://arxiv.org/abs/hep-ph/9310208}{{\ttfamily hep-ph/9310208}}].

\bibitem{Braaten}
E.~Braaten, H.~W. Hammer and G.~P. Lepage, \emph{{Open Effective Field Theories
  from Deeply Inelastic Reactions}},
  \href{https://doi.org/10.1103/PhysRevD.94.056006}{\emph{Phys. Rev.}
  {\bfseries D94} (2016) 056006}
  [\href{https://arxiv.org/abs/1607.02939}{{\ttfamily 1607.02939}}].

\bibitem{Kohn}
W.~Kohn, \emph{{Cyclotron Resonance and de Haas-van Alphen Oscillations of an
  Interacting Electron Gas}},
  \href{https://doi.org/10.1103/PhysRev.123.1242}{\emph{Phys. Rev.} {\bfseries
  123} (1961) 1242}.

\bibitem{Leutwyler}
H.~Leutwyler, \emph{{Nonrelativistic effective Lagrangians}},
  \href{https://doi.org/10.1103/PhysRevD.49.3033}{\emph{Phys. Rev.} {\bfseries
  D49} (1994) 3033} [\href{https://arxiv.org/abs/hep-ph/9311264}{{\ttfamily
  hep-ph/9311264}}].

\bibitem{SpinResonance}
M.~Oshikawa and I.~Affleck, \emph{Electron spin resonance in $s=\frac{1}{2}$
  antiferromagnetic chains},
  \href{https://doi.org/10.1103/PhysRevB.65.134410}{\emph{Phys. Rev. B}
  {\bfseries 65} (2002) 134410}.

\bibitem{Kaplan_KaonCondensate}
D.~B. Kaplan and A.~E. Nelson, \emph{{Strange Goings on in Dense Nucleonic
  Matter}}, \href{https://doi.org/10.1016/0370-2693(86)90331-X}{\emph{Phys.
  Lett.} {\bfseries B175} (1986) 57}.

\bibitem{Son_Kaon1}
D.~T. Son and M.~A. Stephanov, \emph{{QCD at finite isospin density}},
  \href{https://doi.org/10.1103/PhysRevLett.86.592}{\emph{Phys. Rev. Lett.}
  {\bfseries 86} (2001) 592}
  [\href{https://arxiv.org/abs/hep-ph/0005225}{{\ttfamily hep-ph/0005225}}].

\bibitem{Son_Kaon2}
T.~Sch{\"a}fer, D.~T. Son, M.~A. Stephanov, D.~Toublan and J.~J.~M.
  Verbaarschot, \emph{{Kaon condensation and Goldstone's theorem}},
  \href{https://doi.org/10.1016/S0370-2693(01)01265-5}{\emph{Phys. Lett.}
  {\bfseries B522} (2001) 67}
  [\href{https://arxiv.org/abs/hep-ph/0108210}{{\ttfamily hep-ph/0108210}}].

\bibitem{Brown_NeutronStar}
G.~E. Brown, V.~Thorsson, K.~Kubodera and M.~Rho, \emph{{A Novel mechanism for
  kaon condensation in neutron star matter}},
  \href{https://doi.org/10.1016/0370-2693(92)91386-N}{\emph{Phys. Lett.}
  {\bfseries B291} (1992) 355}.

\bibitem{Hellerman}
S.~Hellerman, D.~Orlando, S.~Reffert and M.~Watanabe, \emph{{On the CFT
  Operator Spectrum at Large Global Charge}},
  \href{https://doi.org/10.1007/JHEP12(2015)071}{\emph{JHEP} {\bfseries 12}
  (2015) 071} [\href{https://arxiv.org/abs/1505.01537}{{\ttfamily
  1505.01537}}].

\bibitem{Bern1}
L.~Alvarez-Gaume, O.~Loukas, D.~Orlando and S.~Reffert, \emph{{Compensating
  strong coupling with large charge}},
  \href{https://doi.org/10.1007/JHEP04(2017)059}{\emph{JHEP} {\bfseries 04}
  (2017) 059} [\href{https://arxiv.org/abs/1610.04495}{{\ttfamily
  1610.04495}}].

\bibitem{HellermanO41}
S.~Hellerman, N.~Kobayashi, S.~Maeda and M.~Watanabe, \emph{{A Note on
  Inhomogeneous Ground States at Large Global Charge}},
  \href{https://doi.org/10.1007/JHEP10(2019)038}{\emph{JHEP} {\bfseries 10}
  (2019) 038} [\href{https://arxiv.org/abs/1705.05825}{{\ttfamily
  1705.05825}}].

\bibitem{BootstrapLargeQ}
D.~Jafferis, B.~Mukhametzhanov and A.~Zhiboedov, \emph{{Conformal Bootstrap At
  Large Charge}}, \href{https://doi.org/10.1007/JHEP05(2018)043}{\emph{JHEP}
  {\bfseries 05} (2018) 043}
  [\href{https://arxiv.org/abs/1710.11161}{{\ttfamily 1710.11161}}].

\bibitem{Nicolis:2017eqo}
A.~Nicolis and R.~Penco, \emph{{Mutual Interactions of Phonons, Rotons, and
  Gravity}}, \href{https://doi.org/10.1103/PhysRevB.97.134516}{\emph{Phys. Rev.
  B} {\bfseries 97} (2018) 134516}
  [\href{https://arxiv.org/abs/1705.08914}{{\ttfamily 1705.08914}}].

\bibitem{Brauner1loop}
T.~Brauner, \emph{{Spontaneous symmetry breaking in the linear sigma model at
  finite chemical potential: One-loop corrections}},
  \href{https://doi.org/10.1103/PhysRevD.74.085010}{\emph{Phys. Rev.}
  {\bfseries D74} (2006) 085010}
  [\href{https://arxiv.org/abs/hep-ph/0607102}{{\ttfamily hep-ph/0607102}}].

\bibitem{Luke}
M.~E. Luke and M.~J. Savage, \emph{{Power counting in dimensionally regularized
  NRQCD}}, \href{https://doi.org/10.1103/PhysRevD.57.413}{\emph{Phys. Rev.}
  {\bfseries D57} (1998) 413}
  [\href{https://arxiv.org/abs/hep-ph/9707313}{{\ttfamily hep-ph/9707313}}].

\bibitem{ogievetsky1974nonlinear}
V.~I. Ogievetsky, \emph{Nonlinear realizations of internal and space-time
  symmetries}, {\emph{Proceedings of the Xth Winter School of Theoretical
  Physics in Karpacz} {\bfseries 1} (1974) 117}.

\bibitem{MoninWheel}
L.~V. Delacr{\'e}taz, S.~Endlich, A.~Monin, R.~Penco and F.~Riva,
  \emph{{(Re-)Inventing the Relativistic Wheel: Gravity, Cosets, and Spinning
  Objects}}, \href{https://doi.org/10.1007/JHEP11(2014)008}{\emph{JHEP}
  {\bfseries 11} (2014) 008} [\href{https://arxiv.org/abs/1405.7384}{{\ttfamily
  1405.7384}}].

\bibitem{LowIHC}
I.~Low and A.~V. Manohar, \emph{{Spontaneously broken space-time symmetries and
  Goldstone's theorem}},
  \href{https://doi.org/10.1103/PhysRevLett.88.101602}{\emph{Phys. Rev. Lett.}
  {\bfseries 88} (2002) 101602}
  [\href{https://arxiv.org/abs/hep-th/0110285}{{\ttfamily hep-th/0110285}}].

\bibitem{IvanovIHC}
E.~A. Ivanov and V.~I. Ogievetsky, \emph{{The Inverse Higgs Phenomenon in
  Nonlinear Realizations}},
  \href{https://doi.org/10.1007/BF01028947}{\emph{Teor. Mat. Fiz.} {\bfseries
  25} (1975) 164}.

\bibitem{GuthNRLimit}
M.~H. Namjoo, A.~H. Guth and D.~I. Kaiser, \emph{{Relativistic Corrections to
  Nonrelativistic Effective Field Theories}},
  \href{https://doi.org/10.1103/PhysRevD.98.016011}{\emph{Phys. Rev.}
  {\bfseries D98} (2018) 016011}
  [\href{https://arxiv.org/abs/1712.00445}{{\ttfamily 1712.00445}}].

\bibitem{NRQCD}
G.~T. Bodwin, E.~Braaten and G.~P. Lepage, \emph{{Rigorous QCD analysis of
  inclusive annihilation and production of heavy quarkonium}},
  \href{https://doi.org/10.1103/PhysRevD.55.5853,
  10.1103/PhysRevD.51.1125}{\emph{Phys. Rev.} {\bfseries D51} (1995) 1125}
  [\href{https://arxiv.org/abs/hep-ph/9407339}{{\ttfamily hep-ph/9407339}}].

\bibitem{HoangNRQCD}
A.~H. Hoang, \emph{{Heavy quarkonium dynamics}},
  \href{https://arxiv.org/abs/hep-ph/0204299}{{\ttfamily hep-ph/0204299}}.

\bibitem{RothTasi}
I.~Z. Rothstein, \emph{{TASI lectures on effective field theories}},  2003,
  \href{https://arxiv.org/abs/hep-ph/0308266}{{\ttfamily hep-ph/0308266}}.

\bibitem{Gri1}
H.~W. Griesshammer, \emph{{Threshold expansion and dimensionally regularized
  NRQCD}}, \href{https://doi.org/10.1103/PhysRevD.58.094027}{\emph{Phys. Rev.}
  {\bfseries D58} (1998) 094027}
  [\href{https://arxiv.org/abs/hep-ph/9712467}{{\ttfamily hep-ph/9712467}}].

\bibitem{Gri2}
H.~W. Griesshammer, \emph{{Power counting and Beta function in NRQCD}},
  \href{https://doi.org/10.1016/S0550-3213(99)00325-9}{\emph{Nucl. Phys.}
  {\bfseries B579} (2000) 313}
  [\href{https://arxiv.org/abs/hep-ph/9810235}{{\ttfamily hep-ph/9810235}}].

\bibitem{Caputo}
A.~Caputo, A.~Esposito and A.~D. Polosa, \emph{{Sub-MeV Dark Matter and the
  Goldstone Modes of Superfluid Helium}},
  \href{https://doi.org/10.1103/PhysRevD.100.116007}{\emph{Phys. Rev. D}
  {\bfseries 100} (2019) 116007}
  [\href{https://arxiv.org/abs/1907.10635}{{\ttfamily 1907.10635}}].

\bibitem{vNRQCD}
M.~E. Luke, A.~V. Manohar and I.~Z. Rothstein, \emph{{Renormalization group
  scaling in nonrelativistic QCD}},
  \href{https://doi.org/10.1103/PhysRevD.61.074025}{\emph{Phys. Rev.}
  {\bfseries D61} (2000) 074025}
  [\href{https://arxiv.org/abs/hep-ph/9910209}{{\ttfamily hep-ph/9910209}}].

\bibitem{Weinberg:1990rz}
S.~Weinberg, \emph{{Nuclear forces from chiral Lagrangians}},
  \href{https://doi.org/10.1016/0370-2693(90)90938-3}{\emph{Phys. Lett. B}
  {\bfseries 251} (1990) 288}.

\bibitem{KaplanEFT}
D.~B. Kaplan, \emph{{Five lectures on effective field theory}},  2005,
  \href{https://arxiv.org/abs/nucl-th/0510023}{{\ttfamily nucl-th/0510023}}.

\bibitem{Moroz1}
S.~Moroz, C.~Hoyos, C.~Benzoni and D.~T. Son, \emph{{Effective field theory of
  a vortex lattice in a bosonic superfluid}},
  \href{https://doi.org/10.21468/SciPostPhys.5.4.039}{\emph{SciPost Phys.}
  {\bfseries 5} (2018) 039} [\href{https://arxiv.org/abs/1803.10934}{{\ttfamily
  1803.10934}}].

\bibitem{Son:2005rv}
D.~T. Son and M.~Wingate, \emph{{General coordinate invariance and conformal
  invariance in nonrelativistic physics: Unitary Fermi gas}},
  \href{https://doi.org/10.1016/j.aop.2005.11.001}{\emph{Annals Phys.}
  {\bfseries 321} (2006) 197}
  [\href{https://arxiv.org/abs/cond-mat/0509786}{{\ttfamily
  cond-mat/0509786}}].

\bibitem{RothsteinFL}
I.~Z. Rothstein and P.~Shrivastava, \emph{{Symmetry Obstruction to Fermi Liquid
  Behavior in the Unitary Limit}},
  \href{https://doi.org/10.1103/PhysRevB.99.035101}{\emph{Phys. Rev.}
  {\bfseries B99} (2019) 035101}
  [\href{https://arxiv.org/abs/1712.07797}{{\ttfamily 1712.07797}}].

\bibitem{RiccardoSILH}
G.~F. Giudice, C.~Grojean, A.~Pomarol and R.~Rattazzi, \emph{{The
  Strongly-Interacting Light Higgs}},
  \href{https://doi.org/10.1088/1126-6708/2007/06/045}{\emph{JHEP} {\bfseries
  06} (2007) 045} [\href{https://arxiv.org/abs/hep-ph/0703164}{{\ttfamily
  hep-ph/0703164}}].

\bibitem{IvanovGravity1}
E.~A. Ivanov and J.~Niederle, \emph{{Gauge Formulation of Gravitation Theories.
  1. The Poincare, De Sitter and Conformal Cases}},
  \href{https://doi.org/10.1103/PhysRevD.25.976}{\emph{Phys. Rev.} {\bfseries
  D25} (1982) 976}.

\bibitem{Beneke}
M.~Beneke and V.~A. Smirnov, \emph{{Asymptotic expansion of Feynman integrals
  near threshold}},
  \href{https://doi.org/10.1016/S0550-3213(98)00138-2}{\emph{Nucl. Phys.}
  {\bfseries B522} (1998) 321}
  [\href{https://arxiv.org/abs/hep-ph/9711391}{{\ttfamily hep-ph/9711391}}].

\bibitem{Manohar_ZeroBin}
A.~V. Manohar and I.~W. Stewart, \emph{{The Zero-Bin and Mode Factorization in
  Quantum Field Theory}},
  \href{https://doi.org/10.1103/PhysRevD.76.074002}{\emph{Phys. Rev.}
  {\bfseries D76} (2007) 074002}
  [\href{https://arxiv.org/abs/hep-ph/0605001}{{\ttfamily hep-ph/0605001}}].

\end{thebibliography}\endgroup
\bibliographystyle{JHEP.bst}

\end{document}